\documentclass[trackchanges,twocolumn]{aastex7}
\usepackage{amsmath}

\renewcommand{\today}{}  
\preprint{}

\begin{document}

\title{The Fate of Transonic Shocks around Black Holes and their Future Astrophysical Implications}

\author[orcid=0000-0002-4064-0446,gname=Indu K. ,sname='Dihingia']{Indu  K. Dihingia}
\affiliation{Tsung-Dao Lee Institute, Shanghai Jiao Tong University, 1 Lisuo Road, Shanghai, 201210, People’s Republic of China}
\affiliation{Institute of Fundamental Physics and Quantum Technology,\& School of Physical Science and Technology, Ningbo University, Ningbo, Zhejiang 315211, China}
\email[show]{ikd4638@gmail.com, ikd4638@sjtu.edu.cn}

\author[orcid=0000-0001-8213-646X,gname=Uniyal, sname='Akhil']{Akhil Uniyal} 
\affiliation{Tsung-Dao Lee Institute, Shanghai Jiao Tong University, 1 Lisuo Road, Shanghai, 201210, People’s Republic of China}
\email{akhil_uniyal@sjtu.edu.cn}

\author[orcid=0000-0002-8131-6730,gname=Yosuke, sname='Mizuno']{Yosuke Mizuno}
\affiliation{Tsung-Dao Lee Institute, Shanghai Jiao Tong University, 1 Lisuo Road, Shanghai, 201210, People’s Republic of China}
\affiliation{School of Physics and Astronomy, Shanghai Jiao Tong University, 800 Dongchuan Road, Shanghai, 200240, People’s Republic of China}
\affiliation{Key Laboratory for Particle Physics, Astrophysics and Cosmology (MOE), Shanghai Key Laboratory for Particle Physics and Cosmology, Shanghai Jiao-Tong University,800 Dongchuan Road, Shanghai, 200240, People's Republic of China}
\affiliation{Institut f\"{u}r Theoretische Physik, Goethe-Universit\"{a}t Frankfurt, Max-von-Laue-Strasse 1, D-60438 Frankfurt am Main, Germany}
\email{mizuno@sjtu.edu.cn}
 
\begin{abstract}
Theoretical models have long predicted the existence of shocks in multi-transonic accretion flows onto a black hole, yet their fate under general relativistic simulations has not been fully tested. In this study, we present results from high-resolution two-dimensional general relativistic hydrodynamic (GRHD) and general relativistic magnetohydrodynamic (GRMHD) simulations of low-angular-momentum accretion flows onto Kerr black holes, focusing on the formation of shocks in transonic accretion flow. We demonstrate that for specific combinations of energy and angular momentum, global shock solutions naturally emerge between multiple sonic points. These shocks are sustained in both corotating and counter-rotating cases, and their locations depend on specific energy, angular momentum, and the spin of the black hole which is in good agreement with analytical solutions. In magnetized flows, weak magnetic fields preserve the shock structure, whereas strong fields suppress it, enhancing turbulence and driving powerful, magnetically dominated jets/outflows. The strength and structure of the outflow also depend on a black hole spin and magnetization, with higher black hole spin parameters leading to faster jets. Shock solutions are found only in super-Alfv\'{e}nic regions, where kinetic forces dominate. Our findings provide important insights into the physics of hot corona formation and jet launching in low-angular-momentum accretion systems such as Sgr~A$^*$ (weak jet/outflow) and X-ray binaries.
\end{abstract}

\keywords{\uat{Accretion discs}{14} --- \uat{Black hole physics}{159} --- \uat{High Energy astrophysics}{739} --- \uat{X-ray binaries}{1811}}

\section{Introduction} 
High‐energy X‐ray and gamma‐ray emission from black‐hole accretion systems~\citep{Belloni:2011vs} is now understood to originate very close to the event horizon, in a hot, tenuous, advection‐dominated flow often referred to as the ``corona". In contrast to the cooler, optically thick Keplerian disc--whose thermal multi‐color blackbody spectrum contributes prominently at lower energies~\citep{Shakura:1972te, Abramowicz:1988sp}--this corona is geometrically thicker and quasi‐spherical, and it up‐scatters soft photons into a hard, power‐law tail via inverse Compton processes. Modern spectral modeling, therefore, typically invokes two components: (i) the standard, high‐angular‐momentum disc responsible for the thermal peak, and (ii) a lower‐angular‐momentum, sub‐Keplerian flow that becomes transonic and forms the shock‐heated corona, thereby producing the non‐thermal continuum~\citep{Chakrabarti:1995mx, Molteni:1996qa, Muchotrzebetal1982, Chakrabarti2018}.

Furthermore, because any infalling matter initially moves at subsonic speeds far from the black hole, it must pass through at least one sonic transition to satisfy causality and reach relativistic infall velocities at the horizon~\citep{Bondi1952,Fukue:1984, Matsumoto-etal1984, Fukue1987, Chakrabarti1989, Chakrabarti:1994zq, Chakrabarti:1996cc, Chakrabarti:1996rc, Chakrabarti:1996ef, Chakrabarti:1997et}. In low‐viscosity, inviscid models, the rapidly increasing centrifugal barrier (scaling as 1/$r^3$) near the black hole stalls the flow, leading to the formation of a standing or oscillating shock~\citep{Fukue:1983, Fukue:1983b, Fukue:1983c, Gaffetetal:1983, Chakrabarti1989, Chakrabarti:1993zf, Molteni-etal1994, Chakrabarti:1997hs, Chakrabarti:1998tz, Das:2009ie, Dihingia-etal2019b}. Across this shock front, the gas experiences a sudden compression and heating, creating a localized jump in density and temperature. Downstream of the shock, the flow reaccelerates and becomes supersonic and plunges into the black hole, thus naturally giving rise to the hot, Comptonizing corona that dominates the high‐energy emission and is considered one of the possibilities to form such a hot corona near the black hole.

In theoretical studies of black‐hole accretion, three canonical disc solutions are often employed depending on the mass‐accretion rate: the radiatively inefficient, advection-dominated accretion flow (ADAF)~\citep{Narayan:1994xi, Narayan:1994is, Yuan:2014gma}, the optically thick, geometrically thin “$\alpha$‐disc,”~\citep{Novikov:1973kta, Shakura:1972te}, and the optically thick, radiation-pressure–dominated “slim disc” model~\citep{Abramowicz:1988sp}. Each of these assumes that the infalling gas retains a substantial specific angular momentum, allowing it to orbit the black hole in a flattened disc structure. However, in several astrophysical environments--most notably the Galactic Center source Sgr~A$^*$~\citep{Meliaetal1992, Narayan:1995ic, Ressler:2018yhi}, the progenitors of long gamma-ray bursts~\citep{Fryer:1999mi, Reesetal1988}, and wind-fed high-mass X-ray binaries~\citep{Smith:2001nj, Tauris:2003pf}--the accreting material may possess only minimal angular momentum. Such low-angular-momentum flows cannot form extended Keplerian discs and instead follow nearly radial trajectories, potentially giving rise to distinct shock structures, outflows, and non-thermal emission signatures. Although these quasi-spherical, sub-Keplerian accretion regimes have received less attention than their high-angular-momentum counterparts, understanding their dynamics is crucial for interpreting the energetics, wind production, and jet launching mechanisms in these systems.

\begin{figure}[ht]
    \centering
    \includegraphics[width=\linewidth]{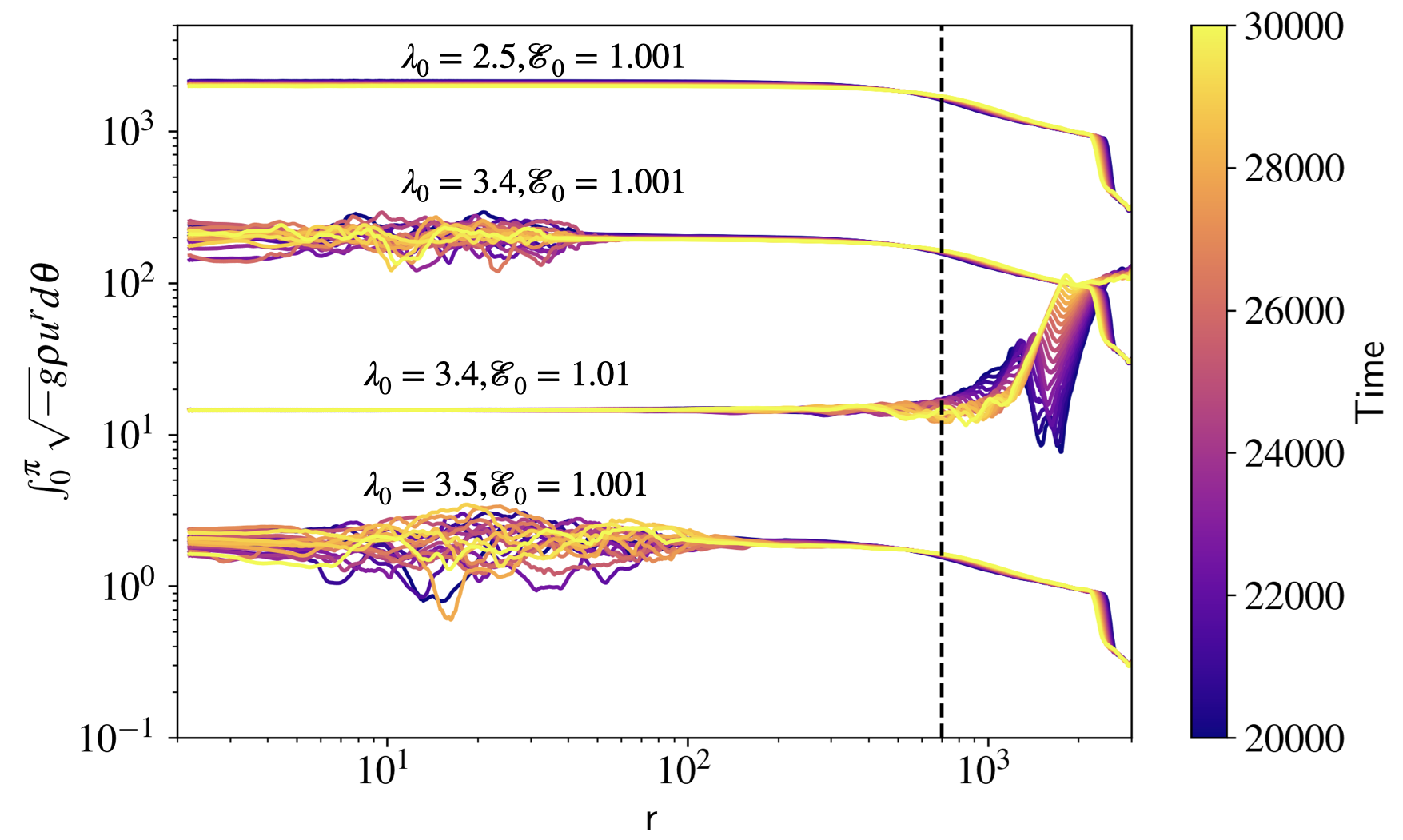}
    \caption{Vertically integrated mass flux ($\int_0^\pi \sqrt{-g}\rho u^r d\theta$ in an arbitrary unit, considering $u^r>0$ as inflow) for the targeted time range ($t=20000-30000\,t_g$) for different simulation models marked on the figure. The dashed vertical line corresponds to $r=700\,r_g$.}
    \label{fig:01A}
\end{figure}

Motivated by these early theoretical investigations of analytical models and extending to recent observational constraints on Sgr~A$^*$, recent studies have explored low angular momentum accretion through both semi-analytic treatments and fully numerical simulations\citep[e.g.,][]{Molteni-etal1994, Ryu-etal1995, Molteni-etal1996a, Molteni-etal1996b, Lanzafame-etal1998, Das-etal2001, Proga-Begelman2003, Chakrabarti-etal2004, Das:2007vf, Das2007, Giri-etal2010, Okuda-Molteni2012, Das-etal2014, Okuda2014, Okuda-Das2015, Kim-etal2017, Okuda-etal2019, Kim-etal2019, Sukova-etal2017, Palit-etal2019, Singh-etal2021, Okuda-etal2022, Garain-Kim2023, Uniyal:2023omh, Dihingia-etal2024,Dihingia-etal2025LA,Mitra-Das2024, Huang-Singh2025,Mao-etal2025}. Several works suggest that, under plausible assumptions, low angular momentum accretion flows can be capable of reproducing Sgr~A$^*$'s observed broadband spectrum and temporal variability \citep{Proga:2002ey, Czerny:2007pb, Moscibrodzka:2007fq, Mach:2018coj}. Numerous studies have demonstrated that relativistic, low-angular-momentum flows can become multi-transonic and sustain both steady and oscillatory shock fronts in $2$D general relativistic hydrodynamic (GRHD) calculations. Extending these efforts into three dimensions, early $3$D GRHD simulations similarly revealed the development of pulsating and radially expanding shocks~\citep{Sukova-etal2017}. More recent high-resolution $3$D GRHD runs, however, have failed to reproduce clear, long-lived standing shocks on a global scale~\citep{Olivares-etal2023}. Despite this, localized density enhancements consistent with transient shock activity have been identified in certain regions of these simulations, even though they do not manifest as coherent, system-wide shock structures. 

In the present work, we focus on the formation and stability of standing shocks within general-relativistic magnetohydrodynamic (GRMHD) flows. While the existence of multiple sonic points and shock formation in low-angular-momentum flows has been established in earlier semi-analytical studies, the behavior of such shocks under the influence of magnetic fields remains elusive. In this work, we extend this classical problem into the GRHD and GRMHD regimes by systematically exploring a wide range of plasma-$\beta$ values and magnetic configurations. We identify the physical conditions, particularly in terms of Alfv\'enic Mach number ${\cal M}_a$, under which global shocks are sustained or suppressed. Our study further establishes a direct connection between shock structure and outflow properties across different black hole spins. These findings provide an alternative framework to study shock-driven jet/outflow formation in weakly magnetized black hole accretion flow.

In the next section, we describe the numerical setup, and in subsequent sections, we discuss our results investigating flow properties and radiative properties. Finally, in section~6, we display our summary and add discussions based on the our findings.

\section{Numerical setup}
We investigate low-angular-momentum accretion flows in a wide range of parameters using 2D ideal GRHD/GRMHD simulations with the \texttt{BHAC} code \citep{Porth-etal2017, Olivares-etal2019} in modified Kerr-Schild coordinates. Simulations use spherical polar coordinates $(r, \theta)$ with logarithmic radial spacing up to $3000\,r_g$, in units where $G = M_{\rm BH} = c = 1$. With this, all length scales and time scales are presented in terms of $r_g=GM_{\rm BH}/c^2$ and $t_g=GM_{\rm BH}/c^3$, respectively. The simulation domain is resolved with an effective resolution of $1024\times 512$ considering two levels of static mesh refinement (SMR) (base resolution $512\times 256$), where maximum SMR levels are employed around the equatorial plane ($\pm 45^\circ$). Additionally, to make the comparison of the study across the different black hole spins, we consider three black hole spin parameters $a_{\rm k}=-0.94, 0, +0.94$. Depending on the spin values, the inner boundary of the simulation domain is set as $r_{\rm b,~in} =0.9r_{\rm H}$, where $r_{\rm H}=(1+\sqrt{1-a^2})~r_g$ is the corresponding event horizon. On the other hand, the outer boundary is set to be $r_{\rm b, out}=3000\,r_g$ for all the simulations.

The initial conditions, viz., four-velocities ($u^\mu$) and density ($\rho$), and pressure $(p)$ are calculated considering $\lambda_0=-u_\phi/u_t={\rm constant}$, and ${\cal E}_0=-hu_t={\rm constant}$, respectively, where $\lambda_0$ and ${\cal E}_0$ are known as the specific angular momentum and specific energy of the fluid element. Detail list of parameters for different simulation models are given in table~\ref{tab:01}. The radial four-velocity ($u^r$) is obtained from semi-analytical solutions for the given $\lambda_0$ and ${\cal E}_0$. With that, all other initial quantities can be calculated. We supplied the initial quantities smoothly connecting the outer edge and event horizon. This ensures that the developed shock discontinuity (if develops) with time evolution is a property of the solution rather than remanent of initial supplied conditions. The methods of getting the explicit expressions can be found in Appendix~\ref{App-A}. 
Earlier axisymmetric (2D) low-angular momentum accretion flow simulations started with nearly vacuum simulation domain and injected supersonic matter from the outer edge (e.g., \cite{Molteni-etal1996, Das-etal2014,Kim-etal2019}). Such injections may always give rise to some kind of discontinuity not necessarily transonic shock. Ideally, we could also inject subsonic matter from the outer edge of the simulation domain $r_{\rm b,~out}=3000\,r_g$ and obtain same solution, but numerical cost will increase in manifolds. Accordingly, we avoided doing so in current study.
We consider the adiabatic approximation to calculate the initial gas pressure considering an index $\Gamma=4/3$, i.e. $p= \kappa \rho^\Gamma$, where $\kappa$ is a constant related to entropy. During time evolution, we use an ideal equation of state, where specific enthalpy is given by $h=1 + \Gamma/(\Gamma-1)~p/\rho$. In the calculation, $g_{\mu\nu}$ corresponds to the metric components of the Kerr black hole in Boyer-Lindquist coordinates.  

As it is standard practice in Eulerian, finite-volume codes, we adopt a floor model to ensure that the simulation code can handle the low-density regions, particularly closer to the black hole and the axis of rotation \citep{Rezzolla-Zanotti2013}. More specifically, we set power law distribution for the floor of rest-mass density and pressure as $\rho_{\rm fl}=10^{-4} r^{-3/2}$ and $p_{\rm fl}=(10^{-6}/3)r^{-5/2}$, respectively. Additionally, we imposed no-inflow conditions at the radial boundaries, which prevent any unphysical accretion from the outer domain. This guarantees that only outflow is permitted across the radial edges of the simulation domain. Along the polar axis, we applied symmetry conditions for scalar variables and radial vector components, while the azimuthal and polar vector components are enforced to be antisymmetric.

\begin{figure*}[ht]
    \centering
    \includegraphics[width=0.8\linewidth]{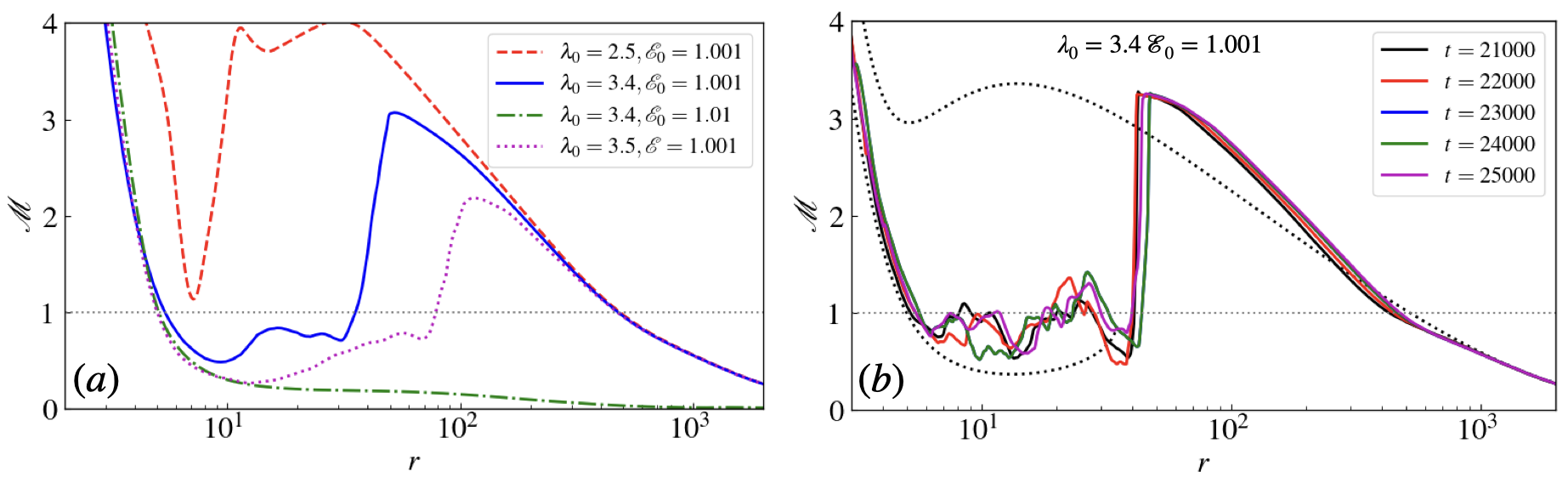}
    \caption{(a) The time and vertically averaged radial Mach number (${\cal M}$) profiles for different pairs of $(\lambda_0, {\cal E}_0)$. (b) Vertically averaged radial Mach number (${\cal M}$) profiles at different simulation times for $(\lambda_0=3.4,{\cal E}_0=1.001$). The horizontal dotted line corresponds to Mach number ${\cal M}=1$. The thick dotted lines in panel (b) correspond to the semi-analytical solutions for the given parameters.}
    \label{fig:01}
\end{figure*}

Note that we evolve the simulations for a very long time, $t=30\,000\,t_g$, so that the impact of initial density distribution can be minimized. During this time, our simulations reach quasi-steady states. All the time-averaged results are shown within $t=20\,000$-$30\,000\,t_g$. 
To justify our choice of simulation time, in Fig.~\ref{fig:01A} we show the radial profiles of the vertically integrated mass flux, $\int_0^\pi \sqrt{-g}\,\rho u^r\,d\theta$ (in arbitrary units), for representative models with $(\lambda_0,{\cal E}_0)=(2.5,1.001), (3.4,1.001), (3.4,1.01)$, and $(3.5,1.001)$, corresponding to model IDs {\tt A0L2P5E1P001}, {\tt A0L3P4E1001}, {\tt A0L3P4E1P01}, and {\tt A0L3P5E1P001} (non-rotating cases), respectively. 
For radial velocity, we follow a sign convention, where $u^r>0$ corresponds to inflow and $u^r<0$ corresponds to outflow.
The profiles exhibit plateau regions, where the accretion rate becomes nearly independent of radius. These plateaus indicate quasi–steady inflow equilibrium. As the simulation evolves, the plateau region expands with time. By $t=30\,000\,t_g$ (yellow curves), a quasi–steady state is achieved up to $r\sim700\,r_g$; this radius is marked by the vertical dashed line in Fig.~\ref{fig:01A}. Beyond $r\gtrsim700\,r_g$, the results cannot be considered reliable. Our primary goal is to study transonic accretion flows, including both inner and outer sonic points. For ${\cal E}_0=1.001$, the typical location of the outer sonic point lies at $r\sim500{-}600\,r_g$ (see Fig.~\ref{fig:01}). From the initial solution, we estimate the infall time, $t_{\rm infall} = \int_{r_{\rm out}}^{r_{\rm H}} \frac{dr}{v(r)},$ from the outer sonic point ($r_{\rm out}$) to the event horizon ($r_{\rm H}$). This gives $t_{\rm infall}\sim 1.0{-}1.2\times 10^4\,t_g$. Thus, our simulation time of $t=30\,000\,t_g$ corresponds to at least three infall timescales, ensuring that the flow within $r\lesssim700\,r_g$ reaches inflow equilibrium. We note that during this period the outermost region ($r\gtrsim2000\,r_g$) gradually empties, due to the very small inflow velocity there ($v\lesssim0.005$). At the end, we think that the best way to simulate such transonic flow for a much longer time would require continuous injection of matter from the outer boundary, which is computationally expensive. Such an injection may alter the accretion flow structure very far from the black hole. Nonetheless, we confirmed that the qualitative transonic nature (sonic point/points, shock formation) remains unaltered from our approach.

\begin{figure*}[ht]
    \centering
    \includegraphics[width=0.9\linewidth]{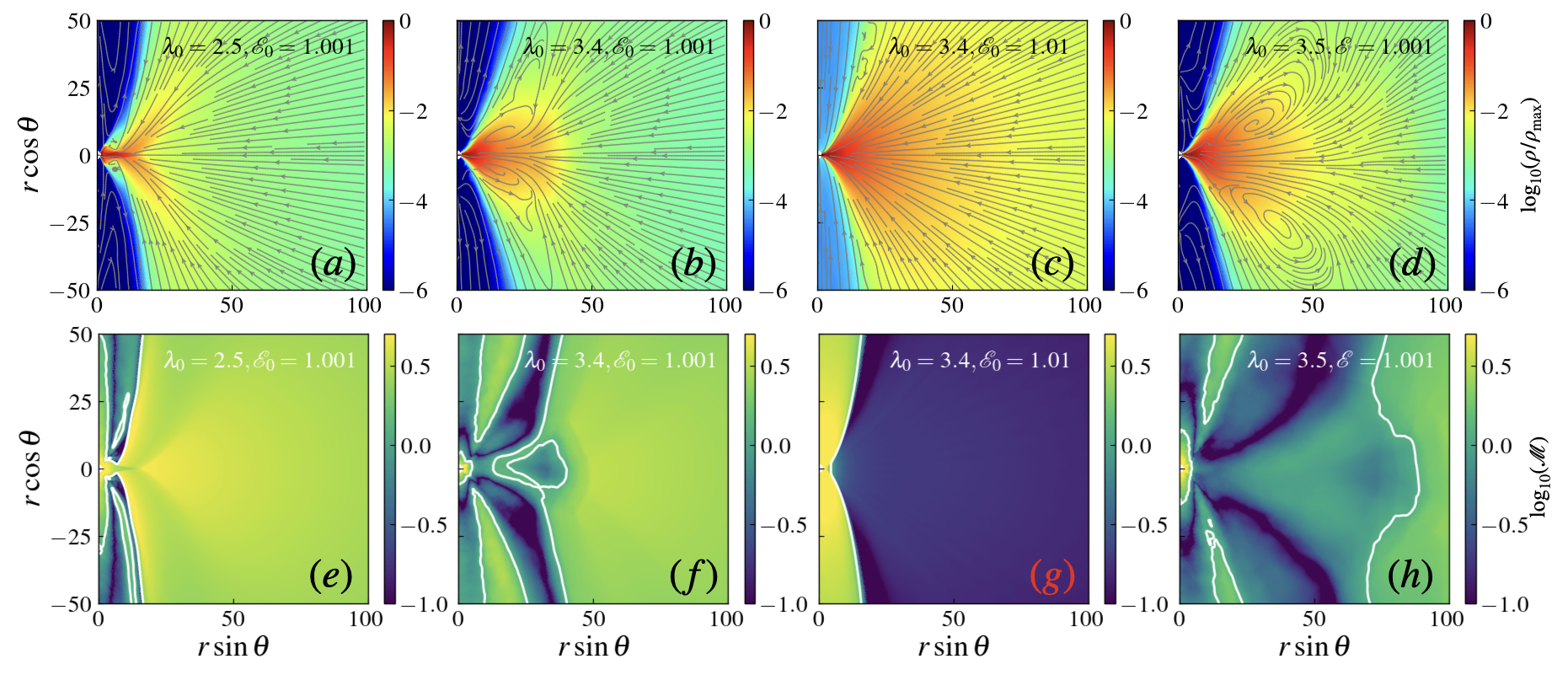}
    \caption{The time-average logarithmic ({\it upper panels}) normalized density ($\log_{\rm 10}(\rho/\rho_{\rm max})$) and ({\it lower panels}) Mach number ($\log_{\rm 10}{\cal M}$) distribution on the poloidal plane for different values of $(\lambda_0, {\cal E}_0)$. The gray lines in the upper panels correspond to the velocity streamlines, and the white lines in the lower panels correspond to the sonic surface ${\cal M}=1$.}
    \label{fig:02}
\end{figure*}

\section{Global Shock solutions}
In this section, we want to study accretion solutions in GRHD by varying specific energy and angular momentum for the Kerr parameter $a_{\rm k}=0.0$. For that, we choose the pairs of specific angular momentum and energy as $(\lambda_0, {\cal E}_0)=(2.5,1.001), (3.4,1.001),(3.4.1.01),$ and $(3.5,1.001)$ with model IDs {\tt A0L2P5E1P001, A0L3P4E1001, A0L3P4E1P01}, and {\tt A0L3P5E1P001}, respectively. The time and vertically averaged radial Mach number (${\cal M}=v^r/a_s$) profiles are shown in Fig.~\ref{fig:01} (a), where the values of $(\lambda_0, {\cal E}_0)$ are marked on the panel. The vertical averaging is performed within $\pm30^\circ$ of the equatorial plane. Note that it is customary to show the Mach number in the corotating frame, where the fluid reaches the black hole horizon with the speed of light. Accordingly, $v^r$ is the radial velocity in the corotating frame and $a_s$ is the relativistic sound speed. To show the location of sonic transitions, we show a horizontal dotted line ${\cal M}=1$. The figure suggests that by choosing the specific energy and angular momentum properly, we could have accretion solutions with only the inner sonic point, only the outer sonic point, and both the critical points. The same results have also been reported before in our earlier study \cite{Dihingia-etal2025}. In addition, the figure shows that for $\lambda_0=3.4,3.5$ and ${\cal E}_0=1.001$ ({\tt A0L3P4E1P001, A0L3P5E1P001}), we observe a sharp transition from supersonic to subsonic flow in between the two sonic points. Due to the time averaging, the sharp transition looks diffused. 
Note that the number of critical points remains preserved with the supplied analytical solution if the sets of parameters have a single sonic point (either outer or inner), e.g., $\lambda_0=2.5, {\cal E}_0=1.001$ and $\lambda_0=3.4, {\cal E}_0=1.01$ ({\tt A0L2P5E1P001, A0L3P4E1P01}). On the contrary, if the parameter sets have multiple sonic points analytically, we observe sharp transitions in the simulations and obtain two sonic points, e.g., $\lambda_0=3.4, {\cal E}_0=1.001$ and $\lambda_0=3.5, {\cal E}_0=1.001$ ({\tt A0L3P4E1P001, A0L3P5E1P001}) although we supplied the solution only passing through the outer sonic point.
To see how the transition looks at different simulation times, in Fig.~\ref{fig:01}b, we show the vertically averaged radial Mach number profiles at different simulation times for $\lambda_0=3.4$ and ${\cal E}_0=1.001$ ({\tt A0L3P4E1P001}). 
In Fig.~\ref{fig:01}b, we also show the semi-analytical solutions for the given parameters by thick dotted lines.
We see that during these simulation times, the transition in between is very sharp around $r_{\rm s}\sim40\,r_g$. Accordingly, we identify it as a shock transition, and such solutions are known as global shock solutions. Moreover, we observe that the location of the shock transition moves far from the black hole with the increase of the angular momentum. 

We see some differences between the semi-analytical and the simulated solution for given sets of parameters ($\lambda_0,{\cal E}_0$), an example can be seen in Fig.~\ref{fig:01}b. This difference is due to the difference in the vertical structure assumptions in the semi-analytical analysis, where a pressure-balanced vertically averaged thin disc ($\alpha/r\ll1$) is considered. However, in the simulations, the accretion disc does not follow the thin-disc assumptions. This accretion flow is ``geometrically thicker'' in nature for low-angular momentum flow; we observe $\alpha/r\gtrsim 1$ (see Appendix~\ref{App-A} and the next section for more detail). Due to a different disc structure than that of the semi-analytical assumption, the obtained shock locations are expected to be slightly different. For example, the semi-analytical shock location for $\lambda_0=3.4$ and ${\cal E}_0=1.001$ is $r_{\rm s}=21.81\,r_g$, but for simulation, we observe shock around $r_{\rm s}\sim40\,r_g$. 
Additionally, we observe some small-scale fluctuations in the Mach number in the post-shock region. They correspond to the hydrodynamic turbulence developed there.

Next, we show the time-average logarithmic normalized density ($\log_{\rm 10}(\rho/\rho_{\rm max})$) and Mach number ($\log_{\rm 10}{\cal M}$) distribution on the poloidal plane for the same pairs $(\lambda_0, {\cal E}_0)$ in the upper and lower panels of Fig.~\ref{fig:02}, respectively. The gray lines in the upper panels correspond to the velocity streamlines, and the white lines in the lower panels correspond to the sonic surface ${\cal M}=1$. In the earlier study \citep{Dihingia-etal2025}, we also reported the variation density profiles depending on the solution types. In the upper panels, we additionally observed that the origin of the outflow streamlines is close to the black hole after the shock transition (2nd and 4th columns). For solutions without shock, most of the streamlines are infalling. We find that some of the outflow emerging from close to the black hole cannot reach infinity. It falls back and creates vortices on both upper and lower hemisphere. 

In the lower panels of Fig.~\ref{fig:02}, we observe that the Mach number is always greater than unity (${\cal M}>1$), i.e., supersonic. Depending on the values of $(\lambda_0, {\cal E}_0)$, we have an additional sonic surface far from the black hole (at the outer sonic surface), which we do not show here (see Fig. \ref{fig:a2}, Appendix \ref{App-B}). In the 2nd and 4th columns, we observe the shock surface, where the flow becomes supersonic to subsonic. Additionally, for these two panels, we see supersonic outflow in the bipolar direction (see the yellowish region surrounding the rotation axis). We confirm that they are outflow, by following the velocity streamlines in the corresponding panels. In summary, we find that by choosing the $(\lambda_0, {\cal E}_0)$ pairs properly, we could seemingly find a global accretion shock solution. 

\section{Variation with Kerr parameters}

\begin{figure*}[ht]
    \centering
    \includegraphics[width=0.9\linewidth]{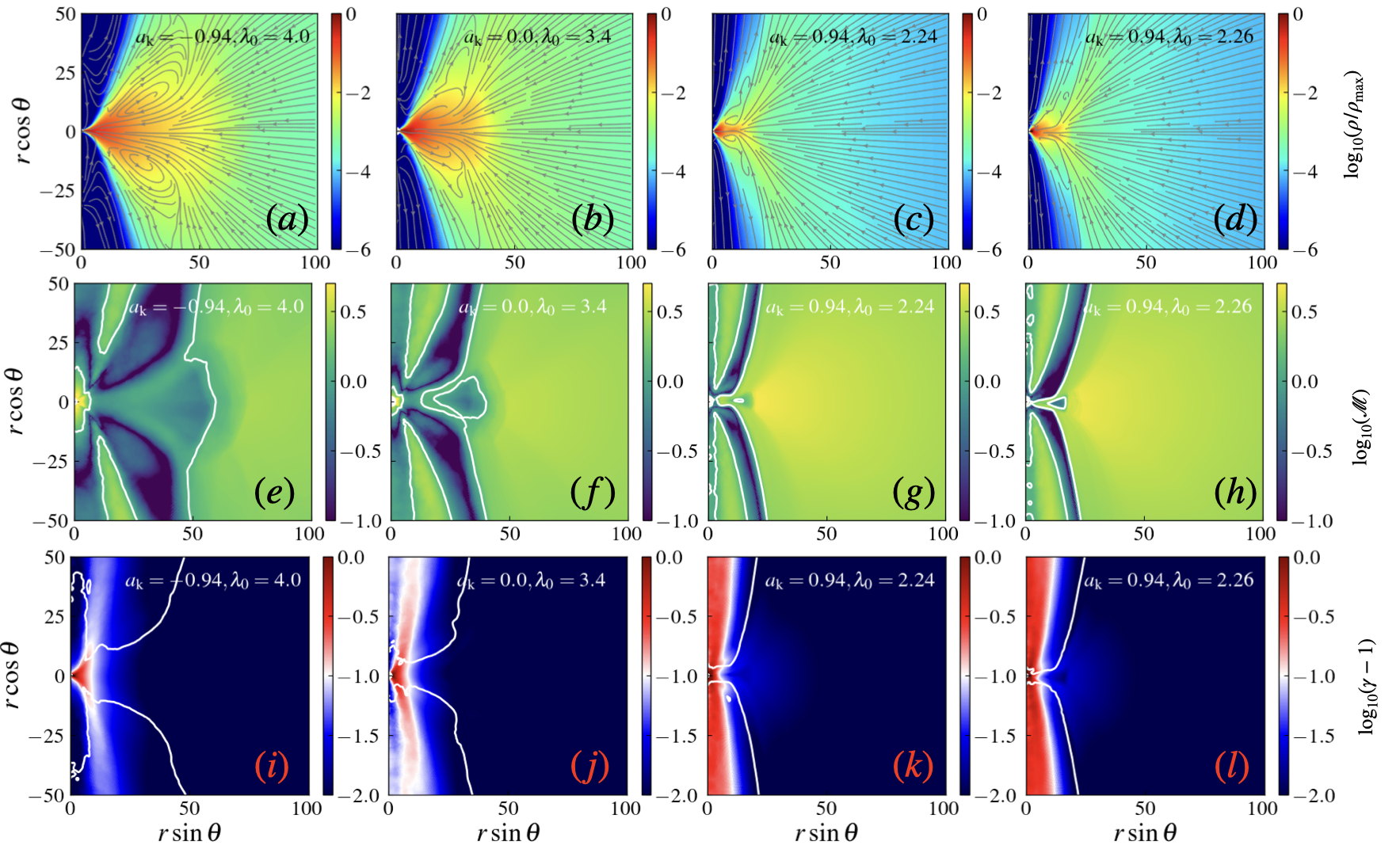}
    \caption{The time-average logarithmic ({\it upper panels}) normalized density ($\log_{\rm 10}(\rho/\rho_{\rm max})$), ({\it middle panels}) Mach number ($\log_{\rm 10}{\cal M}$), and ({\it lower panels}) logarithmic Lorentz factor ($\log_{\rm 10}(\gamma -1)$) distribution on the poloidal plane for different values of Kerr parameters with properly chosen $(\lambda_0, {\cal E}_0=1.001)$. The gray lines in the upper panels correspond to the velocity streamlines, the white lines in the middle panels correspond to the sonic surface ${\cal M}=1$, and the white lines in the lower panels correspond to the outflow surface $\sqrt{-g}\rho u^r=0$.}
    \label{fig:03}
\end{figure*}

\begin{figure}[ht]
    \centering
    \includegraphics[width=0.98\linewidth]{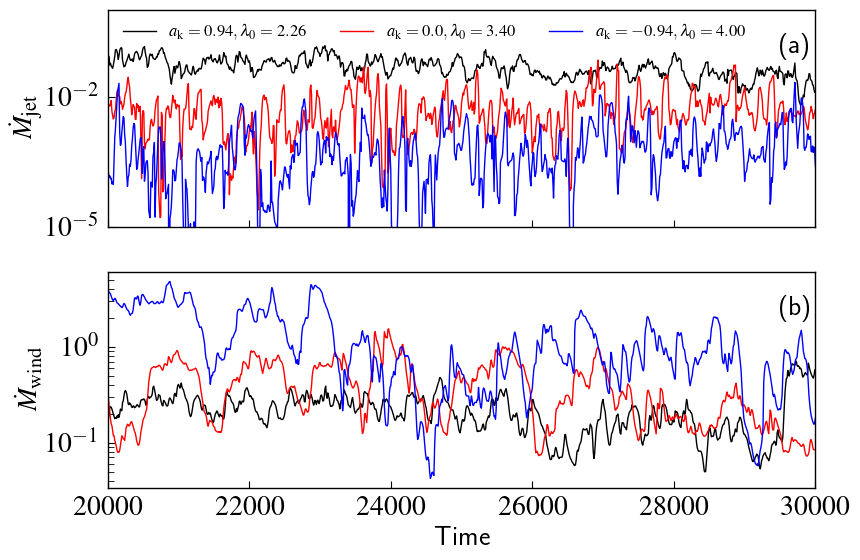}
    \caption{Time evolution of the mass accretion rate through the relativistic jet and non-relativistic wind for different simulation models with different Kerr parameters. See text for more details.}
    \label{fig:03A}
\end{figure}

In the previous section, we found that by choosing $(\lambda_0, {\cal E}_0)$ properly, we could find global accretion shock solutions for the Kerr parameter $a_{\rm k}=0$. In this section, we would like to extend the study for other Kerr parameters. For that in Fig.~\ref{fig:03}, we choose $a_{\rm k}=-0.94$, $0$, and $0.94$ (simulation models {\tt A-L4P0E1P001, A0L3P4E1P001, A+L2P24E1P001, A+L2P26E1P001}), and show the time-average logarithmic ({\it upper panels}) normalized density ($\log_{\rm 10}(\rho/\rho_{\rm max})$), ({\it middle panels}) Mach number ($\log_{\rm 10}{\cal M}$), and ({\it lower panels}) logarithmic Lorentz factor ($\log_{\rm 10}(\gamma -1)$) distribution on the poloidal plane for different values of chosen $\lambda_0$, while specific energy is fixed ${\cal E}_0=1.001$. In the three rows, the lines correspond to velocity streamlines, ${\cal M}=1$ and $\sqrt{-g}\rho u^r=0$, respectively. By observing the 1st and 2nd rows, we find that global accretion shock solutions exist in all the Kerr parameters. Here, we choose the angular momentum such that we have a shock solution by adjusting its values along with the value of the specific energy; the shock surface could form in different locations. In the future, one could do a parameter space survey to find the region in $(\lambda_0, {\cal E}_0)$ space for different Kerr parameters with global accretion shock solutions. Previous semi-analytical calculations suggest that the parameter space moves to lower angular momentum and extended energy range with the increase of Kerr parameters~\citep{Sponholz:1994ek, Chakrabarti:1996cc, Chakrabarti:1996ef, Dihingia-etal2019b}.

In the third row of Fig.~\ref{fig:03}, we see that the Lorentz factor increases with increasing prograde spin ($a_{\rm k}>0$), while it decreases for retrograde spin ($a_{\rm k}<0$). This is not similar to the well-known and well-studied Blandford-Znajek (BZ) process \cite{Blandford-Znajek1977}, where the Lorentz factor and jet power increase with the absolute value of the Kerr parameter~\citep{Komissarov2001, Tchekhovskoy:2009ba, Tchekhovskoy:2011zx, McKinney:2012vh, Tchekhovskoy:2012bg, Liska:2018ayk}. In the BZ process, the jet is launched from the ergosphere; jet properties depend on its characteristics. On the contrary, for low-angular momentum flow, the jet is launched by centrifugal force and thermal pressure. For corotating cases, with the increase of the Kerr parameter, the event horizon radius becomes smaller, which leads to more gravitational compression of the flow, resulting in hotter flow as compared to a case with a lower Kerr parameter. This results in faster jet/outflow from high-spinning cases as compared to low-spinning cases. On the other hand, for the counter-rotating cases, the angular velocity ($\Omega=u^\phi/u^t$) close to the black hole is negative. However, it is a positive in the region far from the black hole. Accordingly, the centrifugal force in the launching region depends monotonically (higher for positive $a_{\rm k}$ and lower for negative $a_{\rm k}$) on the Kerr parameter, resulting in a lower Lorentz factor for the counter-rotating case ($a_{\rm k}=-0.94$, model {\tt A-L4P0E1P001}).

Finally, to demonstrate the dependency of the Kerr parameter on the outflow rate, we show the time evolution of the mass accretion rate through the relativistic jet and non-relativistic wind in Fig.~\ref{fig:03A}. The values of Kerr parameters ($a_{\rm k}$), specific angular momentum ($\lambda_0$) are written on the figure; the specific energy is considered to be ${\cal E}_0=1.001$. Note that these are the same simulation models described earlier in this section. We calculated these quantities by integrating mass flux ($2\pi\int \sqrt{-g}\rho u^r d\theta$) at a radius $r=50\,r_g$. For relativistic jets (panel (a)), we consider outflows with $|\vec{v}|\gtrsim30\%c$, where $g$ is the determinant of the metric. On the other hand, outflow with $|\vec{v}|<30\%c$ is considered wind (panel (b)). 
Since the jet is mildly relativistic, therefore, the contribution of the floor in the mass flux through the jet is negligible. 
Fig.~\ref{fig:03A}a suggests that the relativistic mass flux is highest for the highest Kerr parameter ($a_{\rm k}=0.94$, {\tt A+L2P26E1P001}) and lowest for the lowest Kerr parameter ($a_{\rm k}=-0.94$, {\tt A-L4P0E1P001}). The averaged values of mass flux through jet for $a_{\rm k}=0.94$,$0.0$, and $-0.94$ are given by $\dot{M}_{\rm jet}=5.2\times10^{-2}$, $6.8\times10^{-3}$ and $1.1\times10^{-3}$, respectively. However, the tendency is opposite for non-relativistic wind (Fig.~\ref{fig:03A}b). The average values for mass flux through wind for $a_{\rm k}=0.94$, $0.0$, and $-0.94$ are given by $\dot{M}_{\rm wind}=0.22$, $0.40$, and $1.27$, respectively. Note that these values are expressed in code units, in order to convert them to cgs units, we need to scale them appropriately considering an astrophysical source. 
The values of $\dot{M}_{\rm wind}$ change with the initial angular momentum. To keep the solution type and energy preserved, smaller (counter to corotating) spin values always require higher angular momentum. Therefore, in such scenarios, the observed tendency of the mass flux rate with spin would be unaltered.

\section{Shocks in magnetised flow}\label{sec5}
\begin{figure*}[ht]
    \centering
    \includegraphics[width=0.95\linewidth]{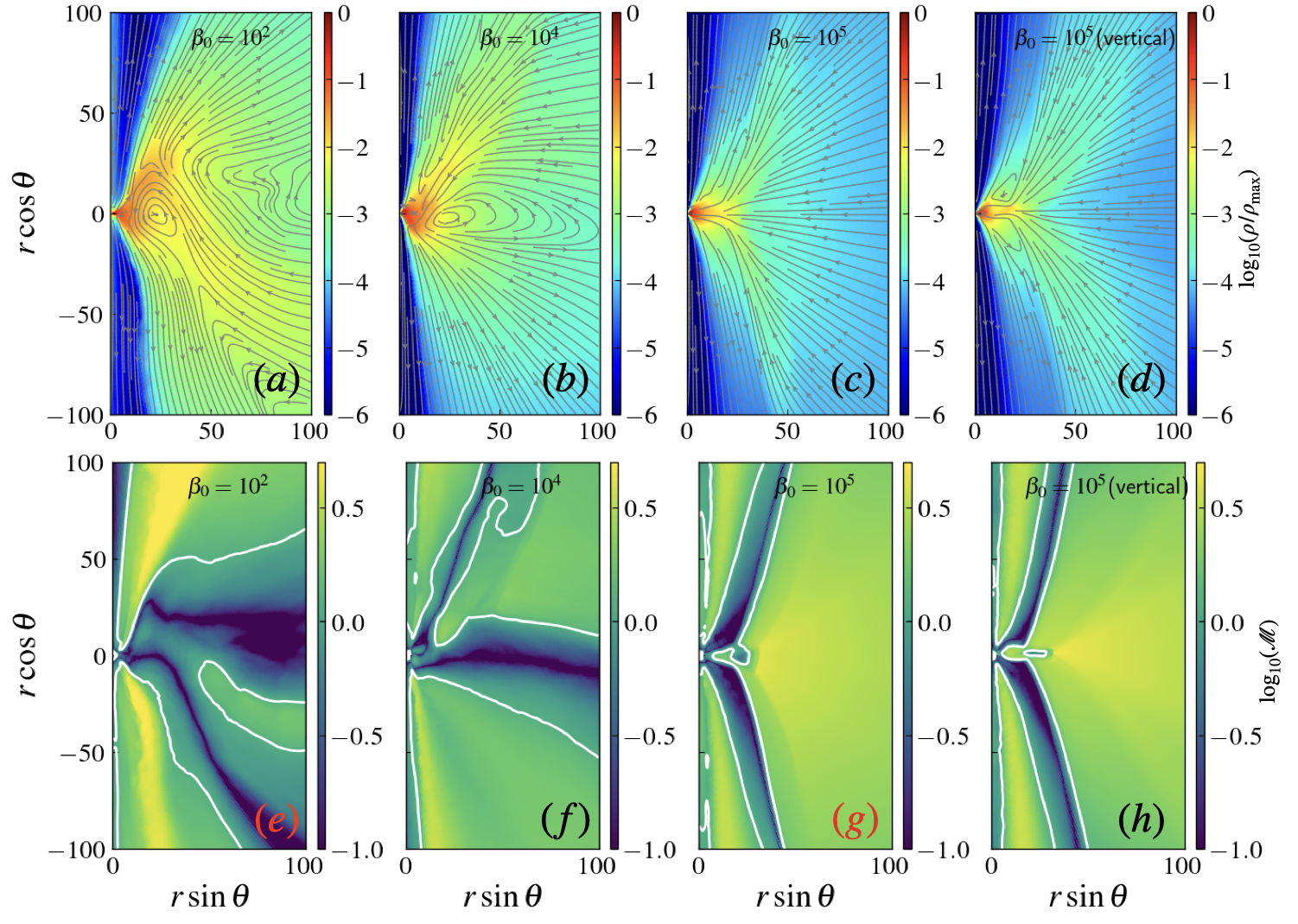}
    \caption{Same as Fig.~\ref{fig:02} but shown in magnetized cases. From left to right column, the cases with inclined magnetic field with different plasma $\beta_0$ ($10^2$ ({\it a,e}), $10^4$ ({\it b,f}), $10^5$ ({\it c,g})), and vertical magnetic field with plasma $\beta_0=10^5$ ({\it d,h}).
    }
    \label{fig:05}
\end{figure*}

In the previous sections, we have discussed low-angular momentum flow without any magnetic fields. In this section, we study the impacts of magnetic field strengths and configurations. 
In order to do that, we consider the case with $a_{\rm k}=0.94$, $\lambda_0=2.26$, and ${\cal E}_0=1.001$. In Fig.~\ref{fig:05}, we show the same quantities 
as Fig.~\ref{fig:02} but for using different magnetic field strengths and configurations. In the first three columns, we show impacts of strengths by choosing initial plasma-$\beta_0=10^2$,$10^4$, and $10^5$ ({\tt A+L2P26E1P001B102, A+L2P26E1P001B104, A+L2P26E1P001B105}), considering inclined magnetic fields. In the fourth row, we show results with the initial vertical magnetic field configuration with the same initial plasma-$\beta_0=10^5$ ({\tt A+L2P26E1P001B105V}). 
In low-angular-momentum, transonic accretion flow, where flow is predominantly radial and the infall time to the black hole is expected to be much shorter than the azimuthal winding time ($t_{\rm in} \sim r/|v_r| \ll t_{\rm w} \sim 1/|{\rm d}\Omega/{\rm d}\ln r|$). Flux freezing implies that $B_{\rm tor}/B_p \sim (\Omega r)/|v_r| \ll 1$. 
This is not true if accretion flow is rotationally supported or predominantly azimuthal.
Therefore, a weak poloidal seed is the reasonable initial condition for predominantly radial accretion flow. Additionally, the differential rotation still winds the initial poloidal field so that a toroidal component develops self-consistently during the run. We initialized the magnetic fields by supplying the vector potential as follows:
\begin{align}
A_\phi(r, \theta) =
\begin{cases}
    \left(r \sin \theta\right)^{3/4} \frac{m^{5/4}}{\left(m^2 + \tan^{-2}(\theta-\pi/2)\right)^{5/8}}\,, & \text{Inclined,} \\
    r \sin\theta, & \text{Vertical.}
\end{cases}
\end{align}
All other components of the vector potential are set to be zero. The inclined magnetic field configuration i.e., vertical field lines are bent radially is set following \cite{Zanni-etal2007,Dihingia-etal2021}. Here we use $m=0.4$. The strength of the magnetic field, is set by supplying the initial plasma-$\beta$ parameter ($\beta_0$). 
Note that with the inclusion of magnetic fields, the steady-state initial solution is also expected to be different. However, initializing such simulations with a consistent initial magnetic field with semi-analytical methods is difficult to achieve due to the unknown configuration of the off-equatorial plane (e.g., \cite{Mitra-etal2022,Mitra-Das2024}). Nonetheless, evolving the simulations sufficiently longer, we achieve a consistent solution with an evolved magnetic field. This may alter shock locations and may eliminate shocks depending on the magnetic field strength for a fixed initial hydrodynamical quantities $(\lambda_0,{\cal E}_0)$. Note that it can be established that (following the same analysis as Fig.~\ref{fig:01A}) the inflow equilibrium radii for these cases are also quite similar to those of the unmagnetized cases ($r\sim700\,r_g$). Only in the strongest magnetic field case (plasma-$\beta_0=10^2$), the equilibrium radius is about $r\sim500\,r_g$. Additionally, the flow becomes more turbulent with the increase in magnetic field strengths within the equilibrium radius.

\begin{figure}[ht]
    \centering
    \includegraphics[width=0.95\linewidth]{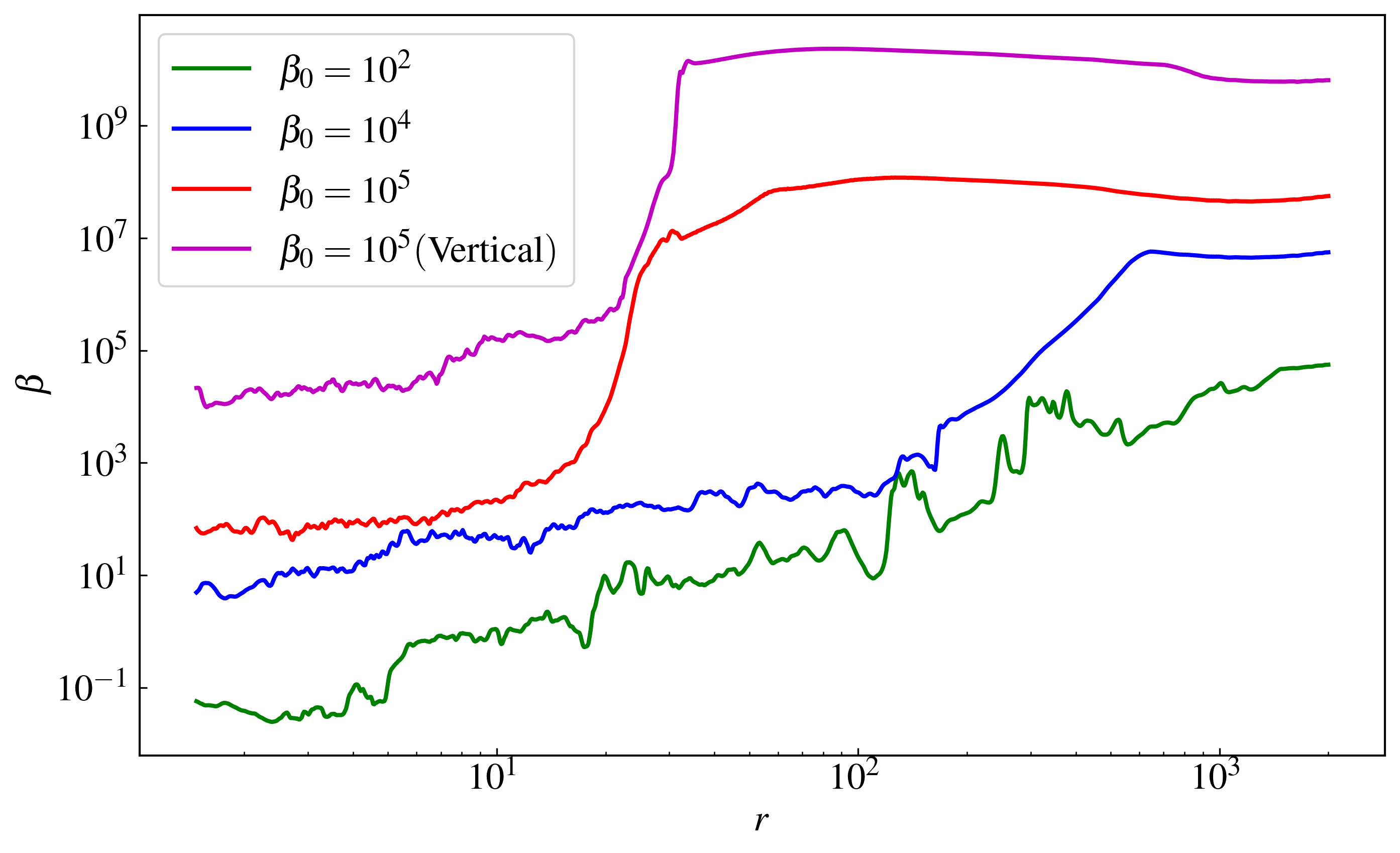}
    \caption{Radial profiles for time-averaged ($t=20000-30000\,t_g$) vertically integrated plasma-$\beta=2p/b^2$ for different initial plasma$\beta$ ($\beta_0$).
    }
    \label{fig:05A}
\end{figure}

To understand the steady-state magnetic field strengths of different simulation cases with different $\beta_0$, i.e., {\tt A+L2P26E1P001B102, A+L2P26E1P001B104, A+L2P26E1P001B105}, and {\tt A+L2P26E1P001B105V}, in Fig.~\ref{fig:05A}, we show radial profiles for time-averaged ($t=20000-30000\,t_g$) vertically integrated plasma-$\beta=2p/b^2$ for these cases. The figure clearly shows an increase in magnetic pressure throughout with the decrease of $\beta_0$. For the cases with $\beta_0=10^5$, the magnitude of plasma-$\beta$ is always $\beta>100$ (gas pressure dominates significantly); thus, results from these cases are expected to be similar to the hydrodynamic cases discussed above. However, for other cases $\beta_0=10^4$ and $\beta_0=10^2$, the value of plasma-$\beta\lesssim10$ near the black hole. Accordingly, we expect a different flow structure for these cases than that of the hydrodynamic cases.

As expected, the first three columns of Fig.~\ref{fig:05} indicate that flow structure remains similar if the magnetic field is weak enough (in this case $\beta_0=10^5$, i.e., {\tt A+L2P26E1P001B105}). In the presence of strong magnetic fields, the enhanced magnetic pressure provides additional support against gravity, effectively slowing down the radially infalling low-angular-momentum matter. This leads to a longer infall timescale, particularly in the inner regions of the accretion flow. 
In low-angular-momentum flows, magnetorotational instability (MRI) can only operate where some differential rotation builds up ~\citep{Balbus:1991ay, Balbus:1998ja, hawley2011assessing, Hawley:2013lga}; a longer infall time allows such shear-supported regions to persist long enough for MRI turbulence to develop. 
Additionally, for a stronger magnetic field, the wavelength of the fastest-growing MRI mode increases. Therefore, it can be resolved with the same numerical resolution. In the weak magnetic field cases ($\beta_0=10^4$ and $10^5$), MRI cannot be resolved unless the field is amplified by advection. The accretion is possible in such cases only due to the low-angular-momentum nature, not due to MRI. Thus, as $\beta_0$ decreases (stronger magnetic field case), the flow becomes more turbulent due to a stronger magnetic field, and the flow near the equatorial plane is always subsonic except very close to the black hole. 
Accordingly, the density distribution becomes more extended with the increase of the magnetic field strength. In such cases, we do not observe any shock formation. For stronger magnetic field cases, we see a large portion of outflowing streamlines as compared to weak magnetic field cases, where streamlines are mostly inflow with a small portion of outflowing streamlines in the post-shock region. Additionally, we observe higher Mach numbers in the bipolar outflow region as compared to lower magnetic field cases, suggesting the development of faster and stronger outflow. The enhanced outflow is primarily due to the increase in magnetocentrifugal acceleration, where torques exerted by the poloidal field lines drive the outflow.

By comparing the 3rd and 4th columns of Fig.~\ref{fig:05}, we examine the effects of different magnetic field configuration: inclined ({\tt A+L2P26E1P001B105}) versus vertical ({\tt A+L2P26E1P001B105V}) initial magnetic field. We compare them for the weak magnetic field limits. Therefore, we observe that the density distribution, velocity streamlines, and the Mach number distribution look quite similar. It suggests that the flow properties of low angular momentum accretion flow remain invariant for different magnetic field configurations, provided the strength is weak. In strong magnetic field cases, the magnetic field dominates the dynamics of the accretion flow (such as in a magnetically arrested disc, MAD). As a result, the flow properties may be drastically different. 

\begin{figure*}[ht]
    \centering
    \includegraphics[width=0.95\linewidth]{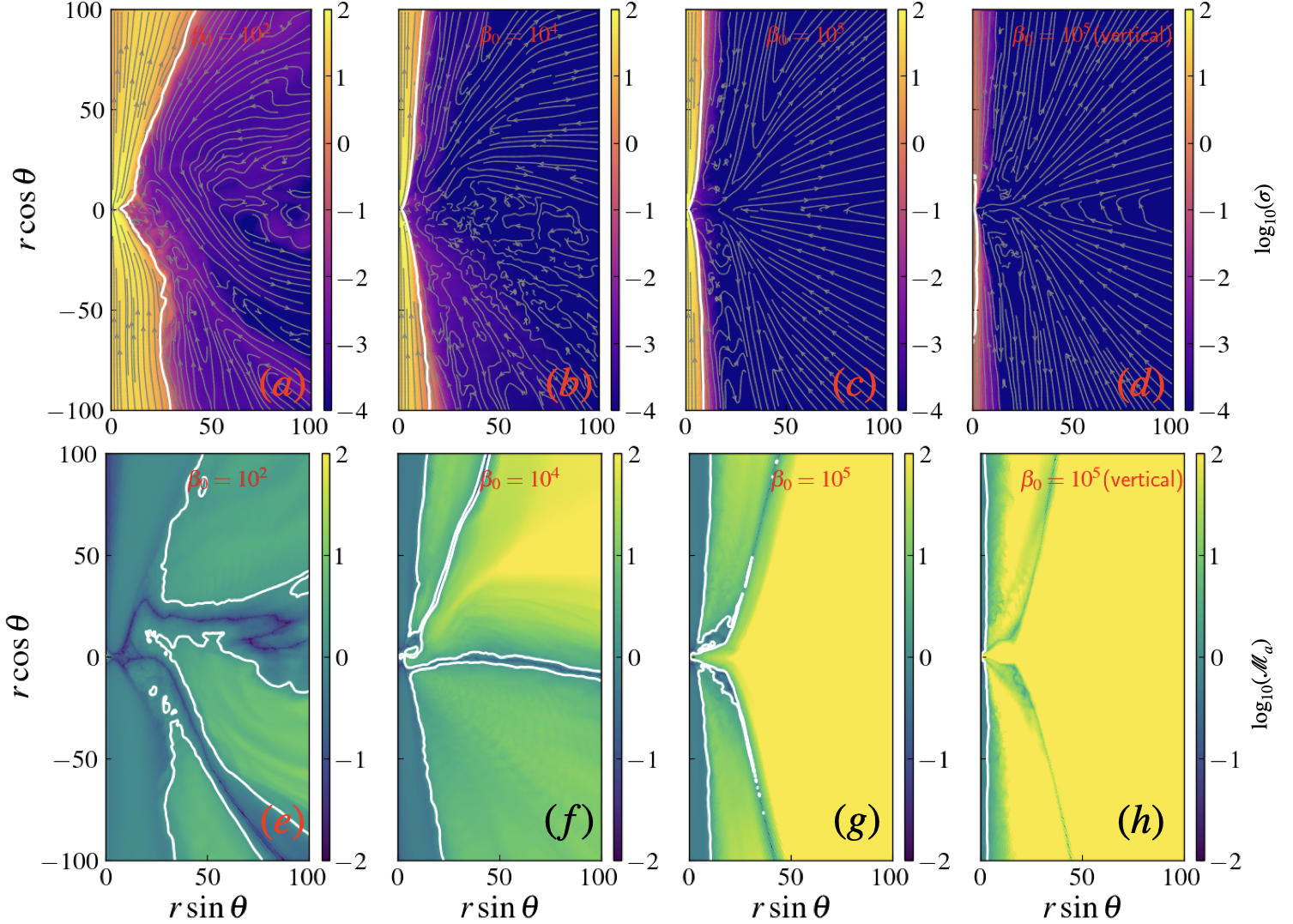}
  \caption{Same as Fig.~\ref{fig:05} but shown
  time-averaged distribution of magnetization ($\sigma=b^2/\rho$ {\it upper}) along with poloidal field lines and the radial Alfv\'enic Mach number (${\cal M}_a=v^r/v_a$, {\it lower}). The solid lines in the upper and lower panels correspond to the boundary of $\sigma=1$ and ${\cal M}_a=1$, respectively.}
    \label{fig:06}
\end{figure*}

To study the magnetized low-angular momentum accretion flow in more detail, in Fig.~\ref{fig:06}, we show the time-averaged distribution of magnetization ($\sigma=b^2/\rho$) along with poloidal field lines and the radial Alfv\'enic Mach number (${\cal M}_a=v^r/v_a$) in the upper and lower panels, respectively, where $v_a$ is the relativistic Alfv\'en velocity $v_a^2=\sigma/(h+\sigma)$. The panels are arranged in exactly the same manner as in Fig.~\ref{fig:05}. The white solid line in the upper and lower panels shows the boundary of $\sigma=1$ and ${\cal M}_a=1$. 
We find that with the increase of magnetic field strength (lowering of $\beta_0$), the magnetization increases, and additionally, the region $\sigma>1$ also monotonically increases with it. For stronger magnetic field cases, the magnetic field lines are very turbulent, outflow from such accretion flow primarily driven by magnetic pressure or magneto-centrifugal force depending on interplay of dominant magnetic field components~\citep{Blandford-Payne1982,Dihingia-etal2021,Dihingia-etal2022JApA}. For weak magnetic field cases, they are organized following velocity streamlines. Suggesting kinetically/gas-dominated flow rather than magnetically dominated flow. This can be further confirmed from the Alfv\'enic Mach number distributions. For weaker magnetic field cases, we observe mostly super-Alfv\'enic (${\cal M}_a>1$) flow except in the bipolar region. This also suggests that magnetic tension and pressure play a minimal role in regulating the dynamics of the bulk flow, except near the polar axis, where magnetic field effects remain significant. On the other hand, with the increase of magnetic field strength (lowering $\beta_0$), the region ${\cal M}_a<1$ increases, and overall the values of Alfv\'enic Mach number decrease, suggesting an increase of the magnetic control over the flow dynamics, particularly due to enhanced magnetic tension and magnetic pressure forces. Moreover, we see a global shock solution only when the flow is super Alfv\'enic (${\cal M}_a\gg1$). Additionally, with the vertical magnetic field case, we also see similar features, as magnetic pressure does not play a significant role in the flow dynamics for weak magnetic field limits.

\begin{figure*}[ht]
    \centering
    \includegraphics[width=0.95\linewidth]{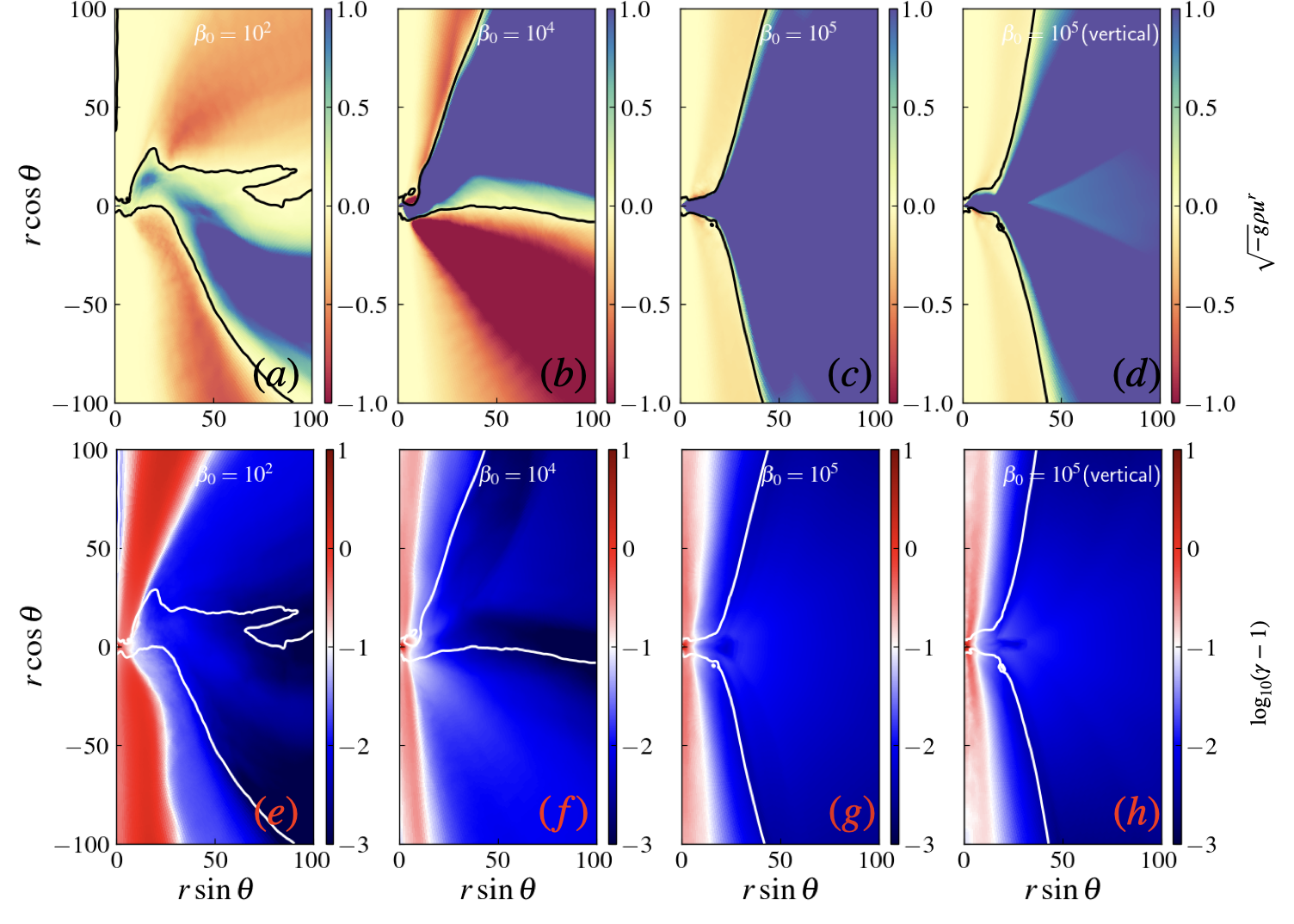}
    \caption{Same as Fig.~\ref{fig:05} but shown the time-averaged distribution of mass flux ($\sqrt{-g}\rho u^r$, {\it upper}) and the Lorentz factor ($\log_{10}(\gamma -1)$, {\it lower}). The solid lines in the upper and lower panels correspond to the boundary of $\sqrt{-g}\rho u^r=0$.}
    \label{fig:07}
\end{figure*}

Next, we would like to study the impacts of magnetic fields on jet/outflow from the accretion flow. In Fig.~\ref{fig:07}, we show the time-averaged distribution of mass flux ($\sqrt{-g}\rho u^r$), and the Lorentz factor ($\log_{10}(\gamma -1)$ in the upper and lower panels, respectively.
The solid lines in the upper and lower panels correspond to the boundary of inflow and outflow ($\sqrt{-g}\rho u^r=0$). For weak magnetic field cases, we observe that the flow is symmetric across the equatorial plane, with bipolar outflow around the rotation axis. With the increase in the magnetic field strength, the inflow breaks its symmetry. 
The stagnation surface ($\sqrt{-g}\rho u^r=0$) marks the shear layer where turbulent disk material interacts with the emerging outflow. MRI-driven turbulence generated close to the black hole is advected outward and seeds fluctuations throughout the inner flow. When this turbulence reaches the stagnation region, its interaction with velocity shear and magnetic-field gradients triggers turbulence. The combined action of MRI-seeded turbulence and shear-driven instabilities naturally amplifies perturbations, producing the large-scale asymmetry observed in the accretion flow. 
We observe qualitatively similar inflow-outflow structure irrespective of the magnetic field configurations in the weak magnetic field limits.
The lower panels suggest that with the increase of the magnetic field strength, the Lorentz factor of the jet increases. In very weak magnetic field cases (irrespective of the magnetic field configuration), the Lorentz factor is of the order of $\gamma\sim2$. On the contrary, in the case of $\beta_0=10^2$ ({\tt A+L2P26E1P001B102}), the value of the Lorentz factor is of the order of $\gamma\sim10$. 
In this case, we observe a strong poloidal magnetic field within the ergosphere, the magnetization in the jet is $\sigma\gg1$ and the normalized time-averaged ($t=20\,000-30\,000\,t_g$) magnetic flux near horizon is of the order of $\Phi/\sqrt{\dot{M}}\sim10$. Such conditions are known to launch a highly relativistic jet via the BZ process \cite[e.g.,][]{Tchekhovskoy:2012bg,Porth-etal2021,Dihingia-etal2021}. For other cases, $\beta_0=10^4$ and $10^5$ ({\tt A+L2P26E1P001B104} and {\tt A+L2P26E1P001B105)}, the normalized time-averaged ($t=20\,000-30\,000\,t_g$) magnetic flux near horizon are of the order of $\Phi/\sqrt{\dot{M}}\sim0.60$ and $0.12$, respectively, which are typically very small to have a jet driven by BZ process.
Ideally, we could increase the magnetic field strength further; this will make the accretion flow close to the limit of the magnetically arrested disc (MAD, $\Phi/\sqrt{\dot{M}}\sim15$) regime. An axisymmetric (2D) simulation, cannot capture the true nature of such MAD accretion flow.
Additionally, we do not expect the flow properties of MAD achieved from traditional torus-based simulations~\citep{Narayan:2003by, Tchekhovskoy:2012bg} or achieved by injecting a strong initial magnetic field in the low-angular momentum flow \citep[e.g.,][]{Kwan-etal2023,Ho-Sang-etal2025} to be different. They cannot be considered as truly a low-angular momentum flow (see Appendix~\ref{App-D} for more details). In such a highly magnetized flow, pressure due to magnetic fields offers an additional barrier to prevent accretion flow from plunging onto the black hole, as in traditional low-angular momentum flow. 
However, this could be different in three-dimensional GRMHD simulations, where the azimuthal interchange instabilities allow spiral stream accretion to form \citep[e.g.,][]{Takasao-etal2019,Porth-etal2021,Begelman-etal2022}.

\begin{figure*}[ht]
    \centering
    \includegraphics[width=0.95\linewidth]{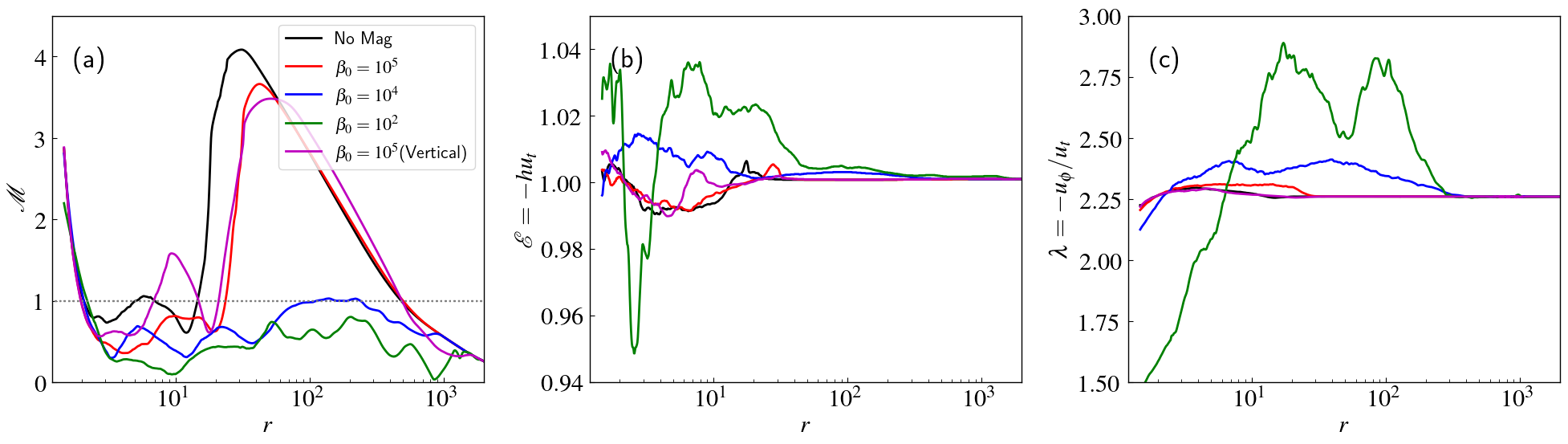}
    \caption{Radial profiles of time and vertically averaged (a) Mach number (${\cal M}$), (b) specific energy (${\cal E}$), and (c) specific angular momentum ($\lambda$) for different magnetic field strengths and configurations.}
    \label{fig:08}
\end{figure*}

Finally, we compare time ($t=20\,000-30\,000\,t_g$) and vertically averaged ($\pm 30^\circ$) radial profiles of (a) Mach number (${\cal M}$), (b) specific energy (${\cal E}=-hu_t$), and (c) specific angular momentum ($\lambda=-u_\phi/u_t$) in the panels of Fig.~\ref{fig:08} for different strengths of magnetic fields and configurations ({\tt A+L2P26E1P001B102, A+L2P26E1P001B104, A+L2P26E1P001B105}, and {\tt A+L2P26E1P001B105V}). As seen in the earlier 2D distributions of variables, radial profiles also suggest similar results. Even with the inclusion of a weak magnetic field (irrespective of configuration), a shock is formed. However, the shock location is slightly outside that of the hydrodynamic case due to the enhancement of the total pressure of the post-shock region by the contribution of magnetic pressure. With the increase of magnetic field strength, we observe that only the inner sonic point is available. It means that shock can not be formed under such conditions.

Panels Fig.~\ref{fig:08}b and Fig.~\ref{fig:08}c suggest that the specific energy and angular momentum are not constant to the input values ${\cal E}_0=1.001, \lambda_0=2.26$, which are expected to be constant if the flow is inviscid and adiabatic. The post-shock region is usually turbulent, irrespective of the magnetic field, due to inflow-outflow instability. Accordingly, we do not observe them to be conserved in the post-shock region. However, their values are conserved in the pre-shock region. 
Additionally, with the inclusion of magnetic fields, MRI-driven turbulence enhances angular-momentum transport, making the magnetized flow more viscous than that of hydrodynamic cases, where we only have viscosity due to numerical dissipation. 
Accordingly, the specific energy increases due to the additional viscous heating in the flow. Similarly, the angular momentum transport increases with the increase of the magnetic field strength. As a result, we see that higher angular momentum outside and lower angular momentum close to the black hole with the increase of the magnetic field strength.

\section{Summary and discussions}
This work explores the fate of global accretion shock solutions by performing a series of GRHD and GRMHD simulations considering different Kerr parameters of the central black hole. We list the comprehensive understanding of our study below. 
\begin{itemize}
  \item[(1)] \textbf{Existence of Global Shock Solutions:}  
  We confirm that global accretion shock solutions form for specific values of the specific energy $({\cal E}_0)$ and angular momentum $(\lambda_0)$, consistent with semi-analytical predictions. These shocks appear as sharp transitions between sonic points. By setting $(\lambda_0,{\cal E}_0)$ properly, such solutions can even be seen in all ranges of Kerr parameters, including both corotating ($a_{\rm k} > 0 $) and counter-rotating ($ a_{\rm k}< 0 $) configurations.

\item[(2)] \textbf{Non-BZ Jet Formation:}  
  Unlike BZ process for jet formation, jet/outflow in our models are launched primarily by centrifugal force and thermal pressure gradients in the hydrodynamic or weakly magnetized cases. This jet formation mechanism would be able to distinguish low-angular-momentum flows from standard disc-jet systems. In this case, the jet/outflow launches from the accretion flow rather than the ergosphere of the black hole.

  \item[(3)] \textbf{Jet Acceleration in Spinning Black Holes:}  
  In the corotating case, a higher black hole spin leads to stronger gravitational compression due to the smaller size of the horizon and strong frame-dragging effect, which produces hotter post-shock flows and faster jets/outflows. The resulting outflows attain higher Lorentz factors with increasing black hole spin. In the counter-rotating case, due to the opposite nature of frame frame-dragging effect with respect to the flow, net centrifugal force decreases, resulting in the formation of weaker and slower jets/outflows.

  \item[(4)] \textbf{Magnetic Field Effects on Shocks:}  
  Very weak magnetic fields ($\beta_0 \geq 10^5$) do not significantly alter the shock structure. However, stronger fields ($\beta_0 \leq 10^2$) suppress shock formation and make the flow turbulent. Thus, the global shock solutions exist only in the limits of high super-Alfv\'{e}nic flow (${\cal M}_a\gg1$). When the Alfv\'{e}nic Mach number drops closer to and below unity (${\cal M}_a\lesssim1$), magnetic forces dominate, suppressing the formation of shock and changing the flow morphology by breaking equatorial plane symmetry of the accretion flow.

  \item[(5)] \textbf{Jet Enhancement by Magnetization:}  
  Magnetically dominated flows exhibit faster and more collimated jets/outflows due to strong magnetic pressure and activation of the BZ process. The Lorentz factor increases from $\gamma \sim 2$ in weak fields to $\gamma \sim 10$ in strong magnetic fields.

  \item[(6)] \textbf{Magnetic Configuration Robustness:}  
  The flow structure is largely insensitive to the magnetic field geometry (inclined vs vertical) in the weak-field limit. The shock and jet properties remain qualitatively similar in both configurations. However, by increasing the magnetic field strength, we may find differences in the flow properties.

  \item[(7)] \textbf{Impact on Flow Energetics and Transport:}  
  Strong magnetic fields enhance angular momentum transport and heating via MRI. As a result, specific energy (${\cal E}=-hu_t$) and angular momentum ($\lambda=-u_\phi/u_t$) are not conserved downstream of the shock, especially in highly magnetized regions.
\end{itemize}

We observed that multi-transonic accretion flows and associated shock structures arise naturally in idealized inviscid and adiabatic models, their realization in actual astrophysical environments around well-known and well-studied astrophysical sources is elusive. Realistic accretion flows are highly dynamic and turbulent, shaped by viscosity, radiative cooling, and magnetic fields, all of which can suppress or destabilize multiple sonic points. Earlier semi-analytic studies also suggest that shock parameter space reduces with viscosity and radiative cooling, and it vanishes in extreme limits \citep[e.g.,][]{Das-Chakrabarti2004,Chakrabarti:2004uy,Kumar-Chattopadhyay2013,Chattopadhyay-Kumar2016,Dihingia-etal2018,Dihingia-etal2019b,Dihingia-etal2020,Kumar:2025erp}. We have also seen the same trend with magnetic field strength. Additionally, the formation of global shock solutions requires fine-tuned combinations of specific angular momentum and energy, conditions that may or may not be generically satisfied for all AGN or Black hole X-ray binaries (BH-XRB) environments. Consequently, even when transient or localized shock-like features arise, they may be short-lived or masked within a turbulent background. It suggests that the presence of long-lived, large-scale shocks is less likely in nature. Thus, these solutions are very good tools to understand idealized accretion flow. Simulation of advanced and complex models invoking a mixture of Keplerian and sub-Keplerian (hot and cold) magnetized accretion flow would, however, be appropriate in most realistic scenarios \citep[e.g.,][]{Esin-etal1997,Chakrabarti-etal2015, Chakrabarti2018}. We plan to do such simulations in the future.

Despite that, we expect such multi-transonic solutions, with shocks and without shocks, can still be useful in certain astrophysical environments, at least in weak magnetic field limits. We list some of such environments (but not all) below:
\begin{itemize}
    \item[(1)] {\bf Radiative inefficient phase of BH-XRBs outburst:}
    Transient BH-XRBs exhibit episodic outbursts lasting from several weeks to months, during which their X-ray luminosity can increase by factors of thousands relative to quiescence. These dramatic flares are generally attributed to a sudden rise in disc viscosity at the “pile‐up” radius, which triggers a rapid inward transport of mass governed by the low angular momentum system~\citep{Chakrabarti1989,Chakrabarti1990, Chakrabarti:1996ns, Chakrabarti:2019clw, Bhowmick:2021lmu}. Accordingly, the GRMHD solutions developed here may offer valuable insights into the outburst behavior of black‐hole X‐ray binaries, and we intend to investigate this application in future work. Future three-dimensional (3D) GRMHD simulations will be able to decipher the physics of observed X-ray polarizations of the black hole sources~\cite[e.g.,][]{Rodriguez-Cavero-etal2023,Steiner-etal2024,Garg-etal2024,Majumder-etal2025}.

    \item[(2)]{\bf Variability in Sgr~A$^*$}
    GRMHD models typically predict rapid, large-amplitude variability in the simulated light curves, which contrasts sharply with the relatively smooth, low-variability emission observed from Sgr~A$^*$~\citep{Murchikova:2021rks, EventHorizonTelescope:2022urf, Murchikova:2022aiz, EventHorizonTelescope:2022ago}. Observations now suggest that Sgr~A$^*$ is primarily fed by the stellar winds of $\sim 30$ massive stars orbiting at parsec scales~\citep{Quataert2004, Cuadra-etal2008, Ressler-etal2018}, resulting in an essentially wind-fed, low-angular-momentum accretion flow rather than a rotation-supported disc. Consequently, extending our sub-Keplerian, low angular momentum GRMHD model for Sgr~A$^*$ may help in understanding the observed variability and will form the basis of our future investigations.
\end{itemize}

Therefore, with the weak magnetization, our multi-transonic GRMHD solutions (with or without shocks) provide a unified framework for modeling low-angular-momentum accretion in a variety of astrophysical settings. Along with the above-mentioned points, our setup can also be used to understand the dynamics of tidal disruption debris, fallback accretion in supernovae and gamma‑ray bursts, and low‑luminosity active galactic nuclei (AGN) fed by Bondi‑type spherical accretion flows. Overall, these solutions capture the interplay between shock formation, flow topology, and radiative inefficiency under weak magnetic fields, offering new insights into diverse accretion phenomena.

Finally, we would like to mention that this study is confined to axisymmetric (2D) GRHD and GRMHD simulations without performing any radiative transfer, thereby restricting the capacity to fully capture three-dimensional turbulence, which could provide direct comparison with observational signatures such as spectra, variability, or polarization. Note that our earlier study hints that radiative properties have direct dependencies on the types of low-angular momentum solutions \citep{Dihingia-etal2025}. It is expected that the low-angular momentum flow is quasi-spherical, and therefore, we expect similar results from the 3D simulations. However, with the increase in magnetic field strengths, such simulations are indeed required. Therefore, the logical progression is to expand the current study with 3D GRMHD simulations and integrate radiative post-processing to evaluate observational significance. Additionally, injecting matter from the outer boundary is a self-consistent way of dealing with transonic accretion flow, which is numerically very expensive, and we do not perform it here. We have confirmed that such an approach only changes the outer part of the accretion flow and helps in obtaining a saturated outer sonic point for the given parameters. It does not alter any qualitative salient features of transonic solutions obtained in the current approach. These efforts are presently in progress and will be communicated in due time.

\begin{acknowledgments}
This research is supported by the National Key R\&D Program of China (Grant No.\,2023YFE0101200), the National Natural Science Foundation of China (Grant No.\,12273022), the Research Fund for Excellent International PhD Students (grant No. W2442004) and the Shanghai Municipality orientation program of Basic Research for International Scientists (Grant No.\,22JC1410600). I.K.D. acknowledges the TDLI postdoctoral fellowship for financial support. The simulations were performed on the TDLI-Astro cluster in Tsung-Dao Lee Institute, Pi2.0, and Siyuan Mark-I clusters in the High-Performance Computing Center at Shanghai Jiao Tong University. This work has made use of NASA's Astrophysics Data System (ADS). We thank the anonymous referee for their careful reading and insightful comments, which have helped to improve the clarity and presentation of this work.
\end{acknowledgments}

\section*{Data Availability}
The simulation data and analysis scripts used in this work are available upon reasonable request. 

\appendix

\section{Initial conditions}\label{App-A}
\begin{figure}
        \centering
        \includegraphics[width=1.0\linewidth]{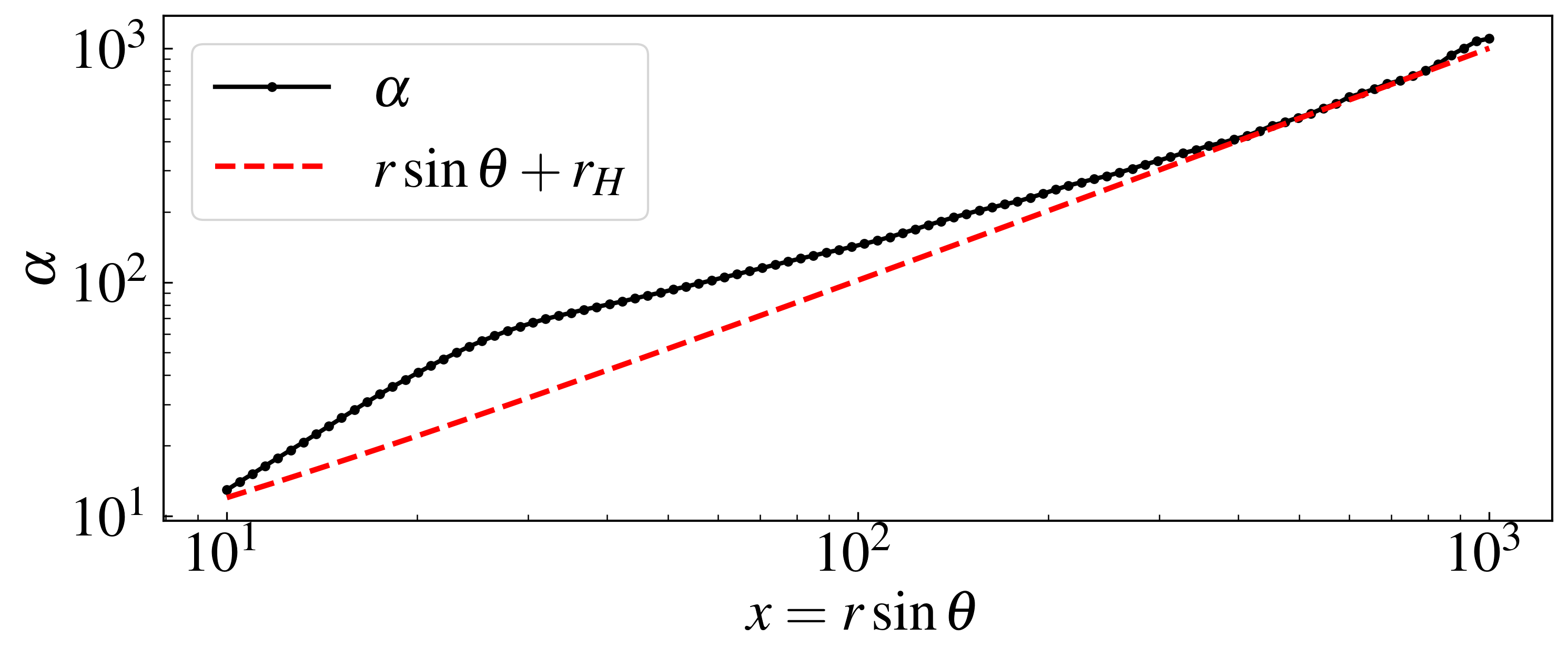}
        \caption{Variation of $\alpha$ along the disc radius (see Appendix \ref{App-A} for more details).}
        \label{fig:a1}
\end{figure}

By simple steps of calculations, with the help of the GRHD equations, the radial derivatives of the radial velocity on the equatorial plane in the corotating frame can be expressed as $dv/dr={\cal N}/{\cal D}$. For explicit expressions of the GRHD equations on the equatorial plane and ${\cal N}$ and ${\cal D}$, please follow \cite{Dihingia-etal2018a,Dihingia-etal2019a}. We solve the equations from the sonic point, where ${\cal N}={\cal D}=0$ simultaneously. Depending on the input parameters, viz., specific energy ${\cal E}_0=-hu_t$, and specific angular momentum $\lambda_0=-u_\phi/u_t$, we can have either one or three sonic points. At the sonic points, we use L'H\^{o}pital's rule to get the two values of the velocity gradient ($dv/dr|_c$). If the sign of both values of $dv/dr|_c$ is real and opposite, such points are known as saddle type or `X-type'. If they are of the same sign, sonic points are known as nodal type. Finally, if the values are imaginary, then the sonic points are known as `O-type'. Here, we only consider `X-type' sonic points with a `-ve' sign for $dv/dr|_c$, which corresponds to the accretion solution. We solve $dv/dr$ and $d\Theta/dr$ (temperature gradient: $\Theta=p/\rho$) starting from the sonic point towards both sides and joining them, we get the full accretion solution connecting the event horizon and infinity. After that, we use the solution $v(r)$ to get all the initial conditions as follows:
\begin{align}
\begin{aligned}
    u^r(r,\theta) &= -g_{rr}^{1/2}\frac{v(r)}{\sqrt{1-v^2(r)}}f(r,\theta), ~~~~
    u^\theta(r,\theta)  = 0,\\
    u^\phi(r,\theta) & = g^{\phi\phi}u_\phi + g^{t\phi}u_t,~~~~
    u^t(r,\theta)  = g^{tt}u_t + g^{t\phi}u_\phi,\\
    \rho(r,\theta) & = \bigg[\frac{\Gamma - 1}{\kappa\Gamma}(h_0 - 1)\bigg]^\frac{1}{\Gamma - 1},~~~~
    p(r,\theta)  = \kappa \rho^\Gamma,\\
\end{aligned}
\end{align}
where
\begin{align}
    h_0 (r,\theta) =  \sqrt{\frac{( 2\lambda_0g^{t\phi} - g^{tt}-\lambda_0^2g^{\phi\phi}){\cal E}_0^2}{1 + g_{rr}(u^r)^2}},~~
    u_t(r,\theta)  = -{\cal E}_0/h_0,~~
     u_\phi(r,\theta)  = -\lambda_0u_t.
\end{align}
Here $f(r,\theta)$ is an assumed function, which models the distribution along the vertical direction. For simplicity, we consider that $u^r$ increases in the vertical direction depending on a scale height, which is given by
\begin{align}
    f(r,\theta) =\exp\left[-\left(\frac{r\cos\theta}{r\sin\theta + r_{\rm H}}\right)^2\right].
\end{align}
The function is motivated by our earlier study with long-term evolution (\cite{Dihingia-etal2025}). We find that the outer part of the accretion flow follows $u^r(r,\theta)=u^r(r,\pi/2)\exp(-(r\cos\theta/\alpha)^2)$. We could roughly set $\alpha=r\sin\theta+r_{\rm H}$, the plot of $\alpha$ and $r\sin\theta + r_{\rm H}$ is shown in Fig.~\ref{fig:a1}. 
The result is demonstrated for $a_{\rm k}=0, \lambda_0=3.4$, and ${\cal E}=1.001$. It is important to choose the parameters appropriately to perform simulations, since the solution topology depends on the values of specific angular momentum ($\lambda_0$) and energy $({\cal E})$. Our choice of parameters is guided by general relativistic semi-analytical calculations by \cite{Dihingia-etal2018a,Dihingia-etal2019a} although a similar range of parameters can be found in some other recent articles \cite[e.g.,][]{Mondal-Basu2020,Tarafdar-etal2021}. We list our simulation models with all relevant parameters in table~\ref{tab:01}:

\begin{table}
\centering
  \begin{tabular}{|c| c | c | c |c | c | c | c |}
    \hline
    Sl. No & ID & Effective resolution & $a_{\rm k}$ & $\lambda_0$ & ${\cal E}_0$ & $\beta_0$ & Field configuration \\ 
    \hline
    1 & \tt{A0L2P5E1P001} & $1024\times 512$ &  $0$ & $2.5$ & $1.001$ & NA & NA\\
    2 & \tt{A0L3P4E1P001} & $1024\times 512$ &  $0$ & $3.4$ & $1.001$ & NA & NA\\
    3 & \tt{A0L3P4E1P01} & $1024\times 512$ &  $0$ & $3.4$ & $1.01$ & NA & NA\\
    4 & \tt{A0L3P5E1P001} & $1024\times 512$ &  $0$ & $3.5$ & $1.001$ & NA & NA\\
    5 & \tt{A-L4P0E1P001} & $1024\times 512$ &  $-0.94$ & $4.0$ & $1.001$ & NA & NA\\
    6 & \tt{A+L2P24E1P001} & $1024\times 512$ &  $0.94$ & $2.24$ & $1.001$ & NA & NA\\
    7 & \tt{A+L2P26E1P001} & $1024\times 512$ &  $0.94$ & $2.26$ & $1.001$ & NA & NA\\
    8 & \tt{A+L2P26E1P001B102} & $1024\times 512$ &  $0.94$ & $2.26$ & $1.001$ & $10^2$ & Inclined\\
    9 & \tt{A+L2P26E1P001B104} & $1024\times 512$ &  $0.94$ & $2.26$ & $1.001$ & $10^4$ & Inclined\\
    10 & \tt{A+L2P26E1P001B105} & $1024\times 512$ &  $0.94$ & $2.26$ & $1.001$ & $10^5$ & Inclined\\
    11 & \tt{A+L2P26E1P001B105V} & $1024\times 512$ &  $0.94$ & $2.26$ & $1.001$ & $10^5$ & Vertical\\
    \hline
  \end{tabular}
\caption{The explicit values of effective resolution, Kerr parameter ($a_{\rm k}$), specific angular momentum ($\lambda_0$), specific energy (${\cal E}_0$), initial plasma-$\beta$ ($\beta_0$), and magnetic field configurations along with corresponding model identifier (ID) for different simulation models.}
\label{tab:01}
\end{table}

\section{Outer sonic surface}\label{App-B}
\begin{figure}
        \centering
        \includegraphics[width=0.9\linewidth]{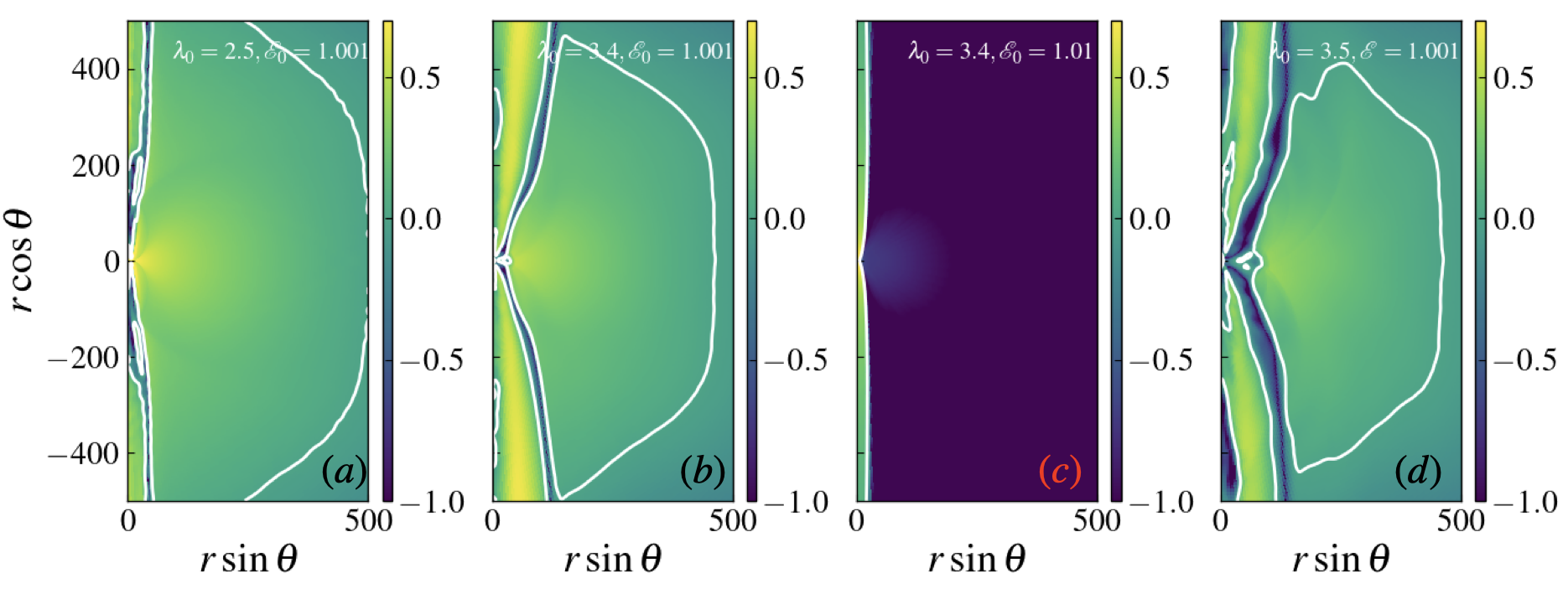}
        \caption{Same as Fig.~\ref{fig:02} but shown larger spatial scale.}
        \label{fig:a2}
\end{figure}

In Fig.~\ref{fig:02} (upper panels), we show the flow properties only within $r<100\,r_g$. However, the outer sonic points in Fig.~\ref{fig:01} are around $r\sim500\,r_g$. To capture the outer sonic surface, in Fig.~\ref{fig:a2}, we show the distribution of Fig.~\ref{fig:02} (upper panels) up to $r<500\,r_g$. The panels show extended outer sonic surfaces for panels (a), (b), and (d). We observe that the outer sonic surface is missing for panel (c), indicating that the solution related to this particular choice of $(\lambda_0, {\cal E}_0)$ does not exhibit an outer sonic point. 

\section{Angular velocity in different black hole spin cases}\label{App-C}
Earlier, we have seen that the jet Lorentz factor increases monotonically with the value (not the absolute value as in the BZ process) of the Kerr parameter. Since the jet/outflow launches from close to the black hole rather than the black hole ergosphere itself. Therefore, the Lorentz factor depends on the gravitational compression as well as the net centrifugal force, which depends on the angular velocity $\Omega=u^\phi/u^t$. In Fig.~\ref{fig:a3}, we show the angular velocity close to the black hole for three different Kerr parameter cases with ${\cal E}_0=1.001$. The value of black hole spin and specific angular momentum is marked on each panel. For the highly spinning case ($a_{\rm k}=0.94$), we observe very high angular velocity around the bipolar region, which is not the case for $a_{\rm k}=-0.94$. In counter-rotating case with $a_{\rm k}=-0.94$, the angular velocity is negative very close to the black hole, which is expected due to the frame-dragging effects. For the zero-spinning case, we observe a higher value of angular velocity than that of the counter-rotating case. The maximum angular velocity region resides off the bipolar region for $a_{\rm k}=0$ (see Fig. \ref{fig:a2}). 
\begin{figure}
        \centering
        \includegraphics[width=0.9\linewidth]{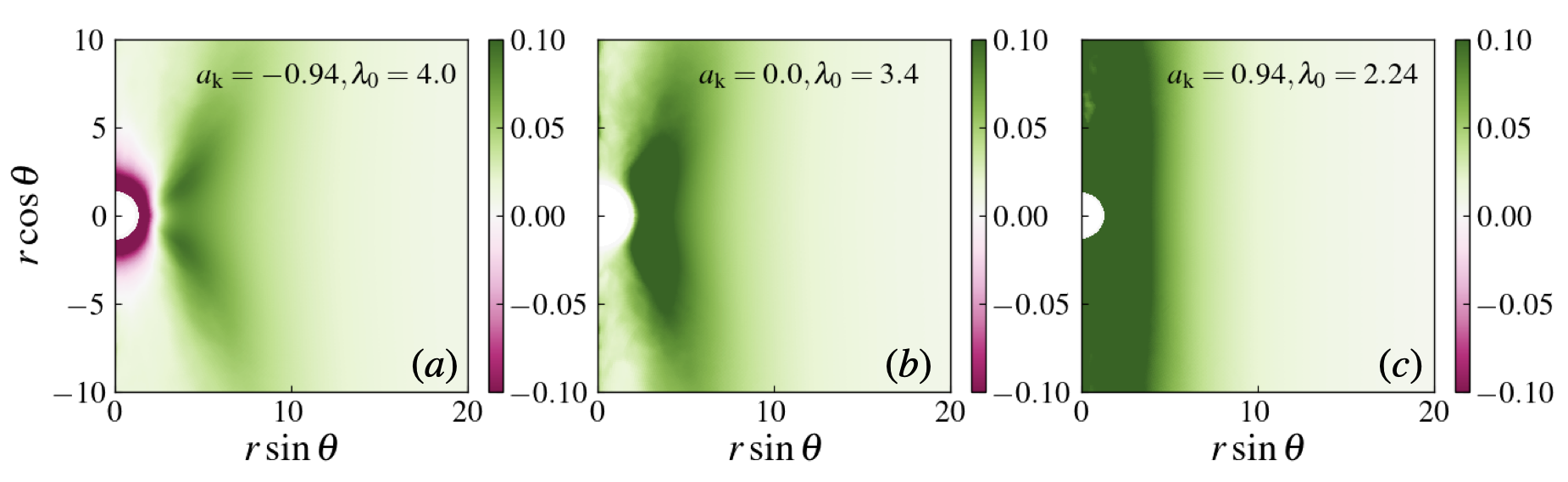}
        \caption{Distribution of angular velocity $\Omega=u^\phi/u_t$ for different simulation models with different Kerr parameters.}
        \label{fig:a3}
\end{figure}

\section{Comparison with hydrostatic equilibrium torus}\label{App-D}
We have seen in Sec.~\ref{sec5} that, with the increase in magnetic field strength, the flow becomes mostly subsonic far from the black hole. In this section, we compare it with Standard and Normal Evolution (SANE) Fishbone-Moncrief torus simulation following \cite{Fishbone-Moncrief1976, Uniyal:2024sdv}. We set $r_{\rm in}=6\,r_g$ and $r_{\rm max}=12\,r_g$ and evolved it up to $t=12000\,t_g$ with a single-loop poloidal magnetic field. We perform a time-averaging with the range of $t=10000-12000\,t_g$ for comparison with our earlier simulation of the case with parameters ${\cal E}=1.001$, $\lambda_0=2.26$, $\beta_0=10^2$, and $a_{\rm k}=0.94$. Fig.~\ref{fig:a3} shows the time-averaged Mach number (${\cal M}$) for low-angular momentum flow with ${\cal E}_0=1.001$, $\lambda_0=2.26$, and $\beta_0=10^2$, (b) for the SANE FM torus. Panels (c), (d), and (e) are showing vertically and time-averaged Mach number, specific energy, and specific angular momentum, respectively. The low angular momentum simulation is shown in black color, with the FM torus simulation in red color. We observe that both results remain very similar near the black hole; however, there are some differences in the outer disc. Furthermore, the transition from subsonic to supersonic in the flow is almost at the same radius, indicating that a low angular momentum flow with a high magnetic field is very much similar to the usual torus settings.
\begin{figure}
        \centering
        \includegraphics[width=0.9\linewidth]{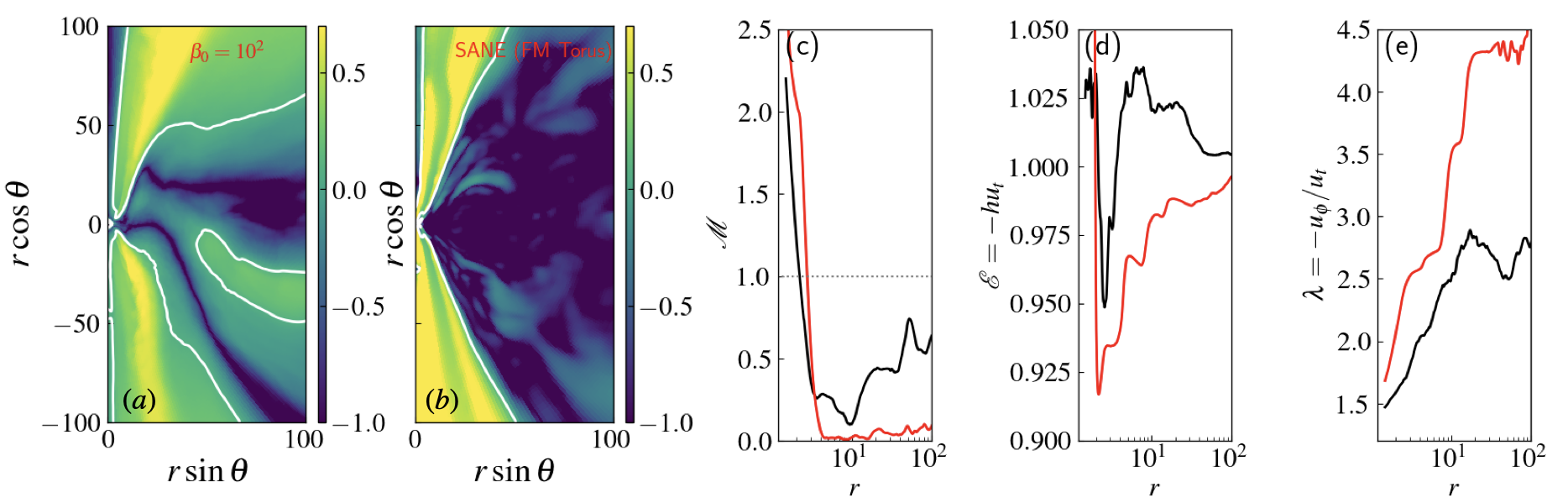}
        \caption{Time-averaged Mach number (${\cal M}$) distribution for (a) low-angular momentum flow with ${\cal E}_0=1.001,\lambda_0=2.26,\beta_0=10^2$ and (b) SANE FM torus. White contours in these two panels correspond to ${\cal M}=1$.
        Panels (c), (d), and (e) show the vertically averaged and time-averaged Mach number, specific energy, and specific angular momentum, respectively, for the FM torus (red) and low-angular momentum (black). The horizontal dotted line in panel (c) corresponds to ${\cal M}=1$.}
        \label{fig:a3}
\end{figure}

\bibliography{sample7}{}

@ARTICLE{Porth-etal2017,
       author = {{Porth}, Oliver and {Olivares}, Hector and {Mizuno}, Yosuke and
         {Younsi}, Ziri and {Rezzolla}, Luciano and {Moscibrodzka}, Monika and
         {Falcke}, Heino and {Kramer}, Michael},
        title = "{The black hole accretion code}",
      journal = {Computational Astrophysics and Cosmology},
         year = 2017,
        month = may,
       volume = {4},
       number = {1},
          eid = {1},
        pages = {1},
          doi = {10.1186/s40668-017-0020-2},
archivePrefix = {arXiv},
       eprint = {1611.09720},
 primaryClass = {gr-qc},
       adsurl = {https://ui.adsabs.harvard.edu/abs/2017ComAC...4....1P}
}

@ARTICLE{Olivares-etal2019,
       author = {{Olivares}, Hector and {Porth}, Oliver and {Davelaar}, Jordy and
         {Most}, Elias R. and {Fromm}, Christian M. and {Mizuno}, Yosuke and
         {Younsi}, Ziri and {Rezzolla}, Luciano},
        title = "{Constrained transport and adaptive mesh refinement in the Black Hole Accretion Code}",
      journal = {\aap},
         year = 2019,
        month = sep,
       volume = {629},
          eid = {A61},
        pages = {A61},
          doi = {10.1051/0004-6361/201935559},
archivePrefix = {arXiv},
       eprint = {1906.10795},
 primaryClass = {astro-ph.HE},
       adsurl = {https://ui.adsabs.harvard.edu/abs/2019A&A...629A..61O}
}

@ARTICLE{Zanni-etal2007,
       author = {{Zanni}, C. and {Ferrari}, A. and {Rosner}, R. and {Bodo}, G. and
         {Massaglia}, S.},
        title = "{MHD simulations of jet acceleration from Keplerian accretion disks. The effects of disk resistivity}",
      journal = {\aap},
         year = 2007,
        month = jul,
       volume = {469},
       number = {3},
        pages = {811-828},
          doi = {10.1051/0004-6361:20066400},
archivePrefix = {arXiv},
       eprint = {astro-ph/0703064},
 primaryClass = {astro-ph},
       adsurl = {https://ui.adsabs.harvard.edu/abs/2007A&A...469..811Z}
}

@ARTICLE{Blandford-Payne1982,
       author = {{Blandford}, R.~D. and {Payne}, D.~G.},
        title = "{Hydromagnetic flows from accretion disks and the production of radio jets.}",
      journal = {\mnras},
         year = 1982,
        month = jun,
       volume = {199},
        pages = {883-903},
          doi = {10.1093/mnras/199.4.883},
       adsurl = {https://ui.adsabs.harvard.edu/abs/1982MNRAS.199..883B}
}

@BOOK{Rezzolla-Zanotti2013,
       author = {{Rezzolla}, Luciano and {Zanotti}, Olindo},
        title = "{Relativistic Hydrodynamics}",
         year = 2013,
       adsurl = {https://ui.adsabs.harvard.edu/abs/2013rehy.book.....R}
}

@ARTICLE{Blandford-Znajek1977,
       author = {{Blandford}, R.~D. and {Znajek}, R.~L.},
        title = "{Electromagnetic extraction of energy from Kerr black holes.}",
      journal = {\mnras},
         year = 1977,
        month = may,
       volume = {179},
        pages = {433-456},
          doi = {10.1093/mnras/179.3.433},
       adsurl = {https://ui.adsabs.harvard.edu/abs/1977MNRAS.179..433B}
}

@ARTICLE{Matsumoto-etal1984,
       author = {{Matsumoto}, R. and {Kato}, S. and {Fukue}, J. and {Okazaki}, A.~T.},
        title = "{Viscous transonic flow around the inner edge of geometrically thin accretion disks}",
      journal = {\pasj},
         year = 1984,
        month = jan,
       volume = {36},
       number = {1},
        pages = {71-85},
       adsurl = {https://ui.adsabs.harvard.edu/abs/1984PASJ...36...71M}
}

@ARTICLE{Dihingia-etal2020,
       author = {{Dihingia}, Indu K. and {Das}, Santabrata and {Prabhakar}, Geethu and {Mand
        al}, Samir},
        title = "{Properties of two-temperature magnetized advective accretion flow around rotating black hole}",
      journal = {\mnras},
         year = 2020,
        month = jun,
       volume = {496},
       number = {3},
        pages = {3043-3059},
          doi = {10.1093/mnras/staa1687},
archivePrefix = {arXiv},
       eprint = {1911.02757},
 primaryClass = {astro-ph.HE},
       adsurl = {https://ui.adsabs.harvard.edu/abs/2020MNRAS.496.3043D}
}

@ARTICLE{Dihingia-etal2018,
       author = {{Dihingia}, Indu K. and {Das}, Santabrata and {Mandal}, Samir},
        title = "{Properties of two-temperature dissipative accretion flow around black holes}",
      journal = {\mnras},
         year = 2018,
        month = apr,
       volume = {475},
       number = {2},
        pages = {2164-2177},
          doi = {10.1093/mnras/stx3269},
archivePrefix = {arXiv},
       eprint = {1712.05534},
 primaryClass = {astro-ph.HE},
       adsurl = {https://ui.adsabs.harvard.edu/abs/2018MNRAS.475.2164D}
}

@ARTICLE{Das2007,
       author = {{Das}, Santabrata},
        title = "{Behaviour of dissipative accretion flows around black holes}",
      journal = {\mnras},
         year = 2007,
        month = apr,
       volume = {376},
       number = {4},
        pages = {1659-1670},
          doi = {10.1111/j.1365-2966.2007.11501.x},
archivePrefix = {arXiv},
       eprint = {astro-ph/0610651},
 primaryClass = {astro-ph},
       adsurl = {https://ui.adsabs.harvard.edu/abs/2007MNRAS.376.1659D}
}

@ARTICLE{Dihingia-etal2019a,
       author = {{Dihingia}, Indu K. and {Das}, Santabrata and {Nandi}, Anuj},
        title = "{Low angular momentum relativistic hot accretion flow around Kerr black holes with variable adiabatic index}",
      journal = {Mon. Not. Roy. Astron. Soc.},
         year = "2019",
        month = "Apr",
       volume = {484},
       number = {3},
        pages = {3209-3218},
          doi = {10.1093/mnras/stz168},
archivePrefix = {arXiv},
       eprint = {1901.04293},
 primaryClass = {astro-ph.HE},
       adsurl = {https://ui.adsabs.harvard.edu/abs/2019MNRAS.484.3209D}
}

@ARTICLE{Dihingia-etal2019b,
       author = {{Dihingia}, Indu K. and {Das}, Santabrata and {Maity}, Debaprasad and {Nandi}, Anuj},
        title = "{Shocks in relativistic viscous accretion flows around Kerr black holes}",
      journal = {Mon. Not. Roy. Astron. Soc.},
         year = "2019",
        month = "Jul",
        pages = {1852},
          doi = {10.1093/mnras/stz1933},
archivePrefix = {arXiv},
       eprint = {1903.02856},
 primaryClass = {astro-ph.HE},
       adsurl = {https://ui.adsabs.harvard.edu/abs/2019MNRAS.tmp.1852D}
}

@ARTICLE{Molteni-etal1996,
       author = {{Molteni}, Diego and {Ryu}, Dongsu and {Chakrabarti}, Sandip K.},
        title = "{Numerical Simulations of Standing Shocks in Accretion Flows around Black Holes: A Comparative Study}",
      journal = {\apj},
         year = 1996,
        month = oct,
       volume = {470},
        pages = {460},
          doi = {10.1086/177877},
archivePrefix = {arXiv},
       eprint = {astro-ph/9605116},
 primaryClass = {astro-ph},
       adsurl = {https://ui.adsabs.harvard.edu/abs/1996ApJ...470..460M}
}

@ARTICLE{Dihingia-etal2021,
       author = {{Dihingia}, Indu K. and {Vaidya}, Bhargav and {Fendt}, Christian},
        title = "{Jets, disc-winds, and oscillations in general relativistic, magnetically driven flows around black hole}",
      journal = {\mnras},
         year = 2021,
        month = aug,
       volume = {505},
       number = {3},
        pages = {3596-3615},
          doi = {10.1093/mnras/stab1512},
archivePrefix = {arXiv},
       eprint = {2105.11468},
 primaryClass = {astro-ph.HE},
       adsurl = {https://ui.adsabs.harvard.edu/abs/2021MNRAS.505.3596D}
}

@ARTICLE{Porth-etal2021,
       author = {{Porth}, O. and {Mizuno}, Y. and {Younsi}, Z. and {Fromm}, C.~M.},
        title = "{Flares in the Galactic Centre - I. Orbiting flux tubes in magnetically arrested black hole accretion discs}",
      journal = {\mnras},
         year = 2021,
        month = apr,
       volume = {502},
       number = {2},
        pages = {2023-2032},
          doi = {10.1093/mnras/stab163},
archivePrefix = {arXiv},
       eprint = {2006.03658},
 primaryClass = {astro-ph.HE},
       adsurl = {https://ui.adsabs.harvard.edu/abs/2021MNRAS.502.2023P}
}

@ARTICLE{Ho-Sang-etal2025,
       author = {{Chan}, Ho-Sang and {Dhang}, Prasun and {Dexter}, Jason and {Begelman}, Mitchell C.},
        title = "{The Impact of Plasma Angular Momentum on Magnetically Arrested Flows and Relativistic Jets in Hot Accretion Flows around Black Holes}",
      journal = {\apj},
         year = 2025,
        month = may,
       volume = {985},
       number = {1},
          eid = {135},
        pages = {135},
          doi = {10.3847/1538-4357/adce7f},
archivePrefix = {arXiv},
       eprint = {2504.15489},
 primaryClass = {astro-ph.HE},
       adsurl = {https://ui.adsabs.harvard.edu/abs/2025ApJ...985..135C}
}

@ARTICLE{Kwan-etal2023,
       author = {{Kwan}, Tom M. and {Dai}, Lixin and {Tchekhovskoy}, Alexander},
        title = "{The Effects of Gas Angular Momentum on the Formation of Magnetically Arrested Disks and the Launching of Powerful Jets}",
      journal = {\apjl},
         year = 2023,
        month = apr,
       volume = {946},
       number = {2},
          eid = {L42},
        pages = {L42},
          doi = {10.3847/2041-8213/acc334},
archivePrefix = {arXiv},
       eprint = {2211.12726},
 primaryClass = {astro-ph.HE},
       adsurl = {https://ui.adsabs.harvard.edu/abs/2023ApJ...946L..42K}
}

@ARTICLE{Singh-etal2021,
       author = {{Singh}, Chandra B. and {Okuda}, Toru and {Aktar}, Ramiz},
        title = "{Effects of resistivity on standing shocks in low angular momentum flows around black holes}",
      journal = {arXiv e-prints},
         year = 2021,
        month = jan,
          eid = {arXiv:2101.05474},
        pages = {arXiv:2101.05474},
archivePrefix = {arXiv},
       eprint = {2101.05474},
 primaryClass = {astro-ph.HE},
       adsurl = {https://ui.adsabs.harvard.edu/abs/2021arXiv210105474S}
}

@INPROCEEDINGS{Chakrabarti-etal2015,
       author = {{Chakrabarti}, S.~K.},
        title = "{Whither TCAF?}",
    booktitle = {Astronomical Society of India Conference Series},
         year = 2015,
       series = {Astronomical Society of India Conference Series},
       volume = {12},
        month = jan,
        pages = {9-16},
archivePrefix = {arXiv},
       eprint = {1509.00565},
 primaryClass = {astro-ph.HE},
       adsurl = {https://ui.adsabs.harvard.edu/abs/2015ASInC..12....9C}
}

@article{Kumar:2025erp,
    author = "Kumar, Amit and Chakrabarti, Sayan and Das, Santabrata",
    title = "{Neutrino-dominated Relativistic Viscous Accretion Flows around Rotating Black Holes with Shocks}",
    eprint = "2501.07080",
    archivePrefix = "arXiv",
    primaryClass = "astro-ph.HE",
    doi = "10.3847/1538-4357/adabc8",
    journal = "Astrophys. J.",
    volume = "980",
    number = "1",
    pages = "68",
    year = "2025"
}

@ARTICLE{Mitra-Das2024,
       author = {{Mitra}, Samik and {Das}, Santabrata},
        title = "{Low-angular-momentum General Relativistic Magnetohydrodynamic Accretion Flows around Rotating Black Holes with Shocks}",
      journal = {\apj},
         year = 2024,
        month = aug,
       volume = {971},
       number = {1},
          eid = {28},
        pages = {28},
          doi = {10.3847/1538-4357/ad55cb},
archivePrefix = {arXiv},
       eprint = {2405.16326},
 primaryClass = {astro-ph.HE},
       adsurl = {https://ui.adsabs.harvard.edu/abs/2024ApJ...971...28M}
}

@ARTICLE{Rodriguez-Cavero-etal2023,
       author = {{Rodriguez Cavero}, Nicole and {Marra}, Lorenzo and {Krawczynski}, Henric and {Dov{\v{c}}iak}, Michal and {Bianchi}, Stefano and {Steiner}, James F. and {Svoboda}, Jiri and {Capitanio}, Fiamma and {Matt}, Giorgio and {Negro}, Michela and {Ingram}, Adam and {Veledina}, Alexandra and {Taverna}, Roberto and {Karas}, Vladimir and {Ursini}, Francesco and {Podgorn{\'y}}, Jakub and {Ratheesh}, Ajay and {Suleimanov}, Valery and {Miku{\v{s}}incov{\'a}}, Romana and {Zane}, Silvia and {Kaaret}, Philip and {Muleri}, Fabio and {Poutanen}, Juri and {Malacaria}, Christian and {Petrucci}, Pierre-Olivier and {Gau}, Ephraim and {Hu}, Kun and {Chun}, Sohee and {Agudo}, Iv{\'a}n and {Antonelli}, Lucio A. and {Bachetti}, Matteo and {Baldini}, Luca and {Baumgartner}, Wayne H. and {Bellazzini}, Ronaldo and {Bongiorno}, Stephen D. and {Bonino}, Raffaella and {Brez}, Alessandro and {Bucciantini}, Niccol{\`o} and {Castellano}, Simone and {Cavazzuti}, Elisabetta and {Chen}, Chien-Ting and {Ciprini}, Stefano and {Costa}, Enrico and {De Rosa}, Alessandra and {Del Monte}, Ettore and {Di Gesu}, Laura and {Di Lalla}, Niccol{\`o} and {Di Marco}, Alessandro and {Donnarumma}, Immacolata and {Doroshenko}, Victor and {Ehlert}, Steven R. and {Enoto}, Teruaki and {Evangelista}, Yuri and {Fabiani}, Sergio and {Ferrazzoli}, Riccardo and {Garc{\'\i}a}, Javier A. and {Gunji}, Shuichi and {Hayashida}, Kiyoshi and {Heyl}, Jeremy and {Iwakiri}, Wataru and {Jorstad}, Svetlana G. and {Kislat}, Fabian and {Kitaguchi}, Takao and {Kolodziejczak}, Jeffery J. and {La Monaca}, Fabio and {Latronico}, Luca and {Liodakis}, Ioannis and {Maldera}, Simone and {Manfreda}, Alberto and {Marin}, Fr{\'e}d{\'e}ric and {Marinucci}, Andrea and {Marscher}, Alan P. and {Marshall}, Herman L. and {Massaro}, Francesco and {Mitsuishi}, Ikuyuki and {Mizuno}, Tsunefumi and {Ng}, Chi-Yung and {O'Dell}, Stephen L. and {Omodei}, Nicola and {Oppedisano}, Chiara and {Papitto}, Alessandro and {Pavlov}, George G. and {Peirson}, Abel L. and {Perri}, Matteo and {Pesce-Rollins}, Melissa and {Pilia}, Maura and {Possenti}, Andrea and {Puccetti}, Simonetta and {Ramsey}, Brian D. and {Rankin}, John and {Roberts}, Oliver J. and {Romani}, Roger W. and {Sgr{\`o}}, Carmelo and {Slane}, Patrick and {Spandre}, Gloria and {Soffitta}, Paolo and {Swartz}, Douglas A. and {Tamagawa}, Toru and {Tavecchio}, Fabrizio and {Tawara}, Yuzuru and {Tennant}, Allyn F. and {Thomas}, Nicholas E. and {Tombesi}, Francesco and {Trois}, Alessio and {Tsygankov}, Sergey S. and {Turolla}, Roberto and {Vink}, Jacco and {Weisskopf}, Martin C. and {Wu}, Kinwah and {Xie}, Fei},
        title = "{The First X-Ray Polarization Observation of the Black Hole X-Ray Binary 4U 1630-47 in the Steep Power-law State}",
      journal = {\apjl},
         year = 2023,
        month = nov,
       volume = {958},
       number = {1},
          eid = {L8},
        pages = {L8},
          doi = {10.3847/2041-8213/acfd2c},
archivePrefix = {arXiv},
       eprint = {2305.10630},
 primaryClass = {astro-ph.HE},
       adsurl = {https://ui.adsabs.harvard.edu/abs/2023ApJ...958L...8R}
}

@ARTICLE{Dihingia-etal2022JApA,
       author = {{Dihingia}, Indu K. and {Vaidya}, Bhargav},
        title = "{Properties of the accretion disc, jet and disc-wind around Kerr black hole}",
      journal = {Journal of Astrophysics and Astronomy},
         year = 2022,
        month = jun,
       volume = {43},
       number = {1},
          eid = {23},
        pages = {23},
          doi = {10.1007/s12036-022-09804-z},
       adsurl = {https://ui.adsabs.harvard.edu/abs/2022JApA...43...23D}
}

@ARTICLE{Majumder-etal2025,
       author = {{Majumder}, Seshadri and {Kushwaha}, Ankur and {Singh}, Swapnil and {Jayasurya}, Kiran M. and {Das}, Santabrata and {Nandi}, Anuj},
        title = "{Probing the accretion geometry of black hole X-ray binaries: A multi-mission spectro-polarimetric and timing study}",
      journal = {arXiv e-prints},
         year = 2025,
        month = jun,
          eid = {arXiv:2506.03774},
        pages = {arXiv:2506.03774},
          doi = {10.48550/arXiv.2506.03774},
archivePrefix = {arXiv},
       eprint = {2506.03774},
 primaryClass = {astro-ph.HE},
       adsurl = {https://ui.adsabs.harvard.edu/abs/2025arXiv250603774M}
}

@ARTICLE{Tarafdar-etal2021,
       author = {{Tarafdar}, Pratik and {Maity}, Susovan and {Das}, Tapas K.},
        title = "{Influence of flow thickness on general relativistic low angular momentum accretion around spinning black holes}",
      journal = {\prd},
         year = 2021,
        month = jan,
       volume = {103},
       number = {2},
          eid = {023023},
        pages = {023023},
          doi = {10.1103/PhysRevD.103.023023},
archivePrefix = {arXiv},
       eprint = {2005.01746},
 primaryClass = {astro-ph.HE},
       adsurl = {https://ui.adsabs.harvard.edu/abs/2021PhRvD.103b3023T}
}

@ARTICLE{Mondal-Basu2020,
       author = {{Mondal}, Soumen and {Basu}, Prasad},
        title = "{On the properties of dissipative shocks in the relativistic accretion flows}",
      journal = {\mnras},
         year = 2020,
        month = sep,
       volume = {497},
       number = {2},
        pages = {2119-2132},
          doi = {10.1093/mnras/staa2035},
       adsurl = {https://ui.adsabs.harvard.edu/abs/2020MNRAS.497.2119M}
}

@ARTICLE{Garg-etal2024,
       author = {{Garg}, Akash and {Rawat}, Divya and {M{\'e}ndez}, Mariano},
        title = "{Unveiling the X-ray polarimetric properties of LMC X-3 with IXPE, NICER, and Swift/XRT}",
      journal = {\mnras},
         year = 2024,
        month = jun,
       volume = {531},
       number = {1},
        pages = {585-591},
          doi = {10.1093/mnras/stae1198},
archivePrefix = {arXiv},
       eprint = {2309.06830},
 primaryClass = {astro-ph.HE},
       adsurl = {https://ui.adsabs.harvard.edu/abs/2024MNRAS.531..585G}
}

@ARTICLE{Steiner-etal2024,
       author = {{Steiner}, James F. and {Nathan}, Edward and {Hu}, Kun and {Krawczynski}, Henric and {Dov{\v{c}}iak}, Michal and {Veledina}, Alexandra and {Muleri}, Fabio and {Svoboda}, Jiri and {Alabarta}, Kevin and {Parra}, Maxime and {Bhargava}, Yash and {Matt}, Giorgio and {Poutanen}, Juri and {Petrucci}, Pierre-Olivier and {Tennant}, Allyn F. and {Baglio}, M. Cristina and {Baldini}, Luca and {Barnier}, Samuel and {Bhattacharyya}, Sudip and {Bianchi}, Stefano and {Brigitte}, Maimouna and {Cabezas}, Mauricio and {Cangemi}, Floriane and {Capitanio}, Fiamma and {Casey}, Jacob and {Rodriguez Cavero}, Nicole and {Castellano}, Simone and {Cavazzuti}, Elisabetta and {Chun}, Sohee and {Churazov}, Eugene and {Costa}, Enrico and {Di Lalla}, Niccol{\`o} and {Di Marco}, Alessandro and {Egron}, Elise and {Ewing}, Melissa and {Fabiani}, Sergio and {Garc{\'\i}a}, Javier A. and {Green}, David A. and {Grinberg}, Victoria and {Hadrava}, Petr and {Ingram}, Adam and {Kaaret}, Philip and {Kislat}, Fabian and {Kitaguchi}, Takao and {Kravtsov}, Vadim and {Kub{\'a}tov{\'a}}, Brankica and {La Monaca}, Fabio and {Latronico}, Luca and {Loktev}, Vladislav and {Malacaria}, Christian and {Marin}, Fr{\'e}d{\'e}ric and {Marinucci}, Andrea and {Maryeva}, Olga and {Mastroserio}, Guglielmo and {Mizuno}, Tsunefumi and {Negro}, Michela and {Omodei}, Nicola and {Podgorn{\'y}}, Jakub and {Rankin}, John and {Ratheesh}, Ajay and {Rhodes}, Lauren and {Russell}, David M. and {{\v{S}}lechta}, Miroslav and {Soffitta}, Paolo and {Spooner}, Sean and {Suleimanov}, Valery and {Tombesi}, Francesco and {Trushkin}, Sergei A. and {Weisskopf}, Martin C. and {Zane}, Silvia and {Zdziarski}, Andrzej A. and {Zhang}, Sixuan and {Zhang}, Wenda and {Zhou}, Menglei and {Agudo}, Iv{\'a}n and {Antonelli}, Lucio A. and {Bachetti}, Matteo and {Baumgartner}, Wayne H. and {Bellazzini}, Ronaldo and {Bongiorno}, Stephen D. and {Bonino}, Raffaella and {Brez}, Alessandro and {Bucciantini}, Niccol{\`o} and {Chen}, Chien-Ting and {Ciprini}, Stefano and {De Rosa}, Alessandra and {Del Monte}, Ettore and {Di Gesu}, Laura and {Donnarumma}, Immacolata and {Doroshenko}, Victor and {Ehlert}, Steven R. and {Enoto}, Teruaki and {Evangelista}, Yuri and {Ferrazzoli}, Riccardo and {Gunji}, Shuichi and {Hayashida}, Kiyoshi and {Heyl}, Jeremy and {Iwakiri}, Wataru and {Jorstad}, Svetlana G. and {Karas}, Vladimir and {Kolodziejczak}, Jeffery J. and {Liodakis}, Ioannis and {Maldera}, Simone and {Manfreda}, Alberto and {Marscher}, Alan P. and {Marshall}, Herman L. and {Massaro}, Francesco and {Mitsuishi}, Ikuyuki and {Ng}, Chi-Yung and {O'Dell}, Stephen L. and {Oppedisano}, Chiara and {Papitto}, Alessandro and {Pavlov}, George G. and {Peirson}, Abel L. and {Perri}, Matteo and {Pesce-Rollins}, Melissa and {Pilia}, Maura and {Possenti}, Andrea and {Puccetti}, Simonetta and {Ramsey}, Brian D. and {Roberts}, Oliver J. and {Romani}, Roger W. and {Sgr{\`o}}, Carmelo and {Slane}, Patrick and {Spandre}, Gloria and {Swartz}, Douglas A. and {Tamagawa}, Toru and {Tavecchio}, Fabrizio and {Taverna}, Roberto and {Tawara}, Yuzuru and {Thomas}, Nicholas E. and {Trois}, Alessio and {Tsygankov}, Sergey S. and {Turolla}, Roberto and {Vink}, Jacco and {Wu}, Kinwah and {Xie}, Fei},
        title = "{An IXPE-led X-Ray Spectropolarimetric Campaign on the Soft State of Cygnus X-1: X-Ray Polarimetric Evidence for Strong Gravitational Lensing}",
      journal = {\apjl},
         year = 2024,
        month = jul,
       volume = {969},
       number = {2},
          eid = {L30},
        pages = {L30},
          doi = {10.3847/2041-8213/ad58e4},
archivePrefix = {arXiv},
       eprint = {2406.12014},
 primaryClass = {astro-ph.HE},
       adsurl = {https://ui.adsabs.harvard.edu/abs/2024ApJ...969L..30S}
}

@ARTICLE{Das-Chakrabarti2004,
       author = {{Das}, Santabrata and {Chakrabarti}, Sandip K.},
        title = "{Properties of Accretion Shocks in Viscous Flows with Cooling Effects}",
      journal = {International Journal of Modern Physics D},
         year = 2004,
        month = jan,
       volume = {13},
       number = {9},
        pages = {1955-1972},
          doi = {10.1142/S0218271804005912},
archivePrefix = {arXiv},
       eprint = {astro-ph/0409664},
 primaryClass = {astro-ph},
       adsurl = {https://ui.adsabs.harvard.edu/abs/2004IJMPD..13.1955D}
}

@ARTICLE{Das-etal2001,
       author = {{Das}, Santabrata and {Chattopadhyay}, Indranil and {Chakrabarti}, Sandip K.},
        title = "{Standing Shocks around Black Holes: An Analytical Study}",
      journal = {\apj},
         year = 2001,
        month = aug,
       volume = {557},
       number = {2},
        pages = {983-989},
          doi = {10.1086/321692},
archivePrefix = {arXiv},
       eprint = {astro-ph/0107046},
 primaryClass = {astro-ph},
       adsurl = {https://ui.adsabs.harvard.edu/abs/2001ApJ...557..983D}
}

@ARTICLE{Chattopadhyay-Kumar2016,
       author = {{Chattopadhyay}, Indranil and {Kumar}, Rajiv},
        title = "{Estimation of mass outflow rates from viscous relativistic accretion discs around black holes}",
      journal = {\mnras},
         year = 2016,
        month = jul,
       volume = {459},
       number = {4},
        pages = {3792-3811},
          doi = {10.1093/mnras/stw876},
archivePrefix = {arXiv},
       eprint = {1605.00752},
 primaryClass = {astro-ph.HE},
       adsurl = {https://ui.adsabs.harvard.edu/abs/2016MNRAS.459.3792C}
}

@ARTICLE{Kumar-Chattopadhyay2013,
       author = {{Kumar}, Rajiv and {Chattopadhyay}, Indranil},
        title = "{Estimation of the mass outflow rates from viscous accretion discs}",
      journal = {\mnras},
         year = 2013,
        month = mar,
       volume = {430},
       number = {1},
        pages = {386-402},
          doi = {10.1093/mnras/sts641},
archivePrefix = {arXiv},
       eprint = {1212.4231},
 primaryClass = {astro-ph.HE},
       adsurl = {https://ui.adsabs.harvard.edu/abs/2013MNRAS.430..386K}
}

@ARTICLE{Das-etal2014,
       author = {{Das}, Santabrata and {Chattopadhyay}, Indranil and {Nandi}, Anuj and {Molteni}, D.},
        title = "{Periodic mass loss from viscous accretion flows around black holes}",
      journal = {\mnras},
         year = 2014,
        month = jul,
       volume = {442},
       number = {1},
        pages = {251-258},
          doi = {10.1093/mnras/stu864},
archivePrefix = {arXiv},
       eprint = {1405.4415},
 primaryClass = {astro-ph.HE},
       adsurl = {https://ui.adsabs.harvard.edu/abs/2014MNRAS.442..251D}
}

@ARTICLE{Dihingia-etal2018a,
       author = {{Dihingia}, Indu K. and {Das}, Santabrata and {Maity}, Debaprasad and {Chakrabarti}, Sayan},
        title = "{Limitations of the pseudo-Newtonian approach in studying the accretion flow around a Kerr black hole}",
      journal = {\prd},
         year = 2018,
        month = oct,
       volume = {98},
       number = {8},
          eid = {083004},
        pages = {083004},
          doi = {10.1103/PhysRevD.98.083004},
archivePrefix = {arXiv},
       eprint = {1806.08481},
 primaryClass = {astro-ph.HE},
       adsurl = {https://ui.adsabs.harvard.edu/abs/2018PhRvD..98h3004D}
}

@ARTICLE{Chakrabarti1989,
       author = {{Chakrabarti}, Sandip K.},
        title = "{Standing Rankine-Hugoniot Shocks in the Hybrid Model Flows of the Black Hole Accretion and Winds}",
      journal = {\apj},
         year = 1989,
        month = dec,
       volume = {347},
        pages = {365},
          doi = {10.1086/168125},
       adsurl = {https://ui.adsabs.harvard.edu/abs/1989ApJ...347..365C}
}

@ARTICLE{Sukova-etal2017,
       author = {{Sukov{\'a}}, P. and {Charzy{\'n}ski}, S. and {Janiuk}, A.},
        title = "{Shocks in the relativistic transonic accretion with low angular momentum}",
      journal = {\mnras},
         year = 2017,
        month = dec,
       volume = {472},
       number = {4},
        pages = {4327-4342},
          doi = {10.1093/mnras/stx2254},
archivePrefix = {arXiv},
       eprint = {1709.01824},
 primaryClass = {astro-ph.HE},
       adsurl = {https://ui.adsabs.harvard.edu/abs/2017MNRAS.472.4327S}
}

@ARTICLE{Esin-etal1997,
       author = {{Esin}, Ann A. and {McClintock}, Jeffrey E. and {Narayan}, Ramesh},
        title = "{Advection-Dominated Accretion and the Spectral States of Black Hole X-Ray Binaries: Application to Nova Muscae 1991}",
      journal = {\apj},
         year = 1997,
        month = nov,
       volume = {489},
       number = {2},
        pages = {865-889},
          doi = {10.1086/304829},
archivePrefix = {arXiv},
       eprint = {astro-ph/9705237},
 primaryClass = {astro-ph},
       adsurl = {https://ui.adsabs.harvard.edu/abs/1997ApJ...489..865E}
}

@INPROCEEDINGS{Chakrabarti2018,
       author = {{Chakrabarti}, Sandip K.},
        title = "{Study of accretion processes around black holes becomes `Science': Tell tale observational signatures of two component advective flows}",
    booktitle = {Fourteenth Marcel Grossmann Meeting - MG14},
         year = 2018,
       editor = {{Bianchi}, Massimo and {Jansen}, Robert T. and {Ruffini}, Remo},
        month = jan,
        pages = {369-384},
          doi = {10.1142/9789813226609_0020},
       adsurl = {https://ui.adsabs.harvard.edu/abs/2018mgm..conf..369C}
}

@ARTICLE{Begelman-etal2022,
       author = {{Begelman}, Mitchell C. and {Scepi}, Nicolas and {Dexter}, Jason},
        title = "{What really makes an accretion disc MAD}",
      journal = {\mnras},
         year = 2022,
        month = apr,
       volume = {511},
       number = {2},
        pages = {2040-2051},
          doi = {10.1093/mnras/stab3790},
archivePrefix = {arXiv},
       eprint = {2111.02439},
 primaryClass = {astro-ph.HE},
       adsurl = {https://ui.adsabs.harvard.edu/abs/2022MNRAS.511.2040B}
}

@ARTICLE{Mitra-etal2022,
       author = {{Mitra}, Samik and {Maity}, Debaprasad and {Dihingia}, Indu Kalpa and {Das}, Santabrata},
        title = "{Study of general relativistic magnetohydrodynamic accretion flow around black holes}",
      journal = {\mnras},
         year = 2022,
        month = nov,
       volume = {516},
       number = {4},
        pages = {5092-5109},
          doi = {10.1093/mnras/stac2431},
archivePrefix = {arXiv},
       eprint = {2204.01412},
 primaryClass = {astro-ph.HE},
       adsurl = {https://ui.adsabs.harvard.edu/abs/2022MNRAS.516.5092M}
}

@ARTICLE{Fukue1987,
       author = {{Fukue}, Jun},
        title = "{Transonic disk accretion revisited}",
      journal = {\pasj},
         year = 1987,
        month = jan,
       volume = {39},
       number = {2},
        pages = {309-327},
       adsurl = {https://ui.adsabs.harvard.edu/abs/1987PASJ...39..309F}
}

@ARTICLE{Olivares-etal2023,
       author = {{Olivares}, H{\'e}ctor R. and {Mo{\'s}cibrodzka}, Monika A. and {Porth}, Oliver},
        title = "{General relativistic hydrodynamic simulations of perturbed transonic accretion}",
      journal = {\aap},
         year = 2023,
        month = oct,
       volume = {678},
          eid = {A141},
        pages = {A141},
          doi = {10.1051/0004-6361/202346010},
archivePrefix = {arXiv},
       eprint = {2301.12020},
 primaryClass = {astro-ph.HE},
       adsurl = {https://ui.adsabs.harvard.edu/abs/2023A&A...678A.141O}
}

@ARTICLE{Kim-etal2017,
       author = {{Kim}, Jinho and {Garain}, Sudip K. and {Balsara}, Dinshaw S. and {Chakrabarti}, Sandip K.},
        title = "{General relativistic numerical simulation of sub-Keplerian transonic accretion flows on to black holes: Schwarzschild space-time}",
      journal = {\mnras},
         year = 2017,
        month = nov,
       volume = {472},
       number = {1},
        pages = {542-549},
          doi = {10.1093/mnras/stx1986},
archivePrefix = {arXiv},
       eprint = {1707.09856},
 primaryClass = {astro-ph.HE},
       adsurl = {https://ui.adsabs.harvard.edu/abs/2017MNRAS.472..542K}
}

@ARTICLE{Kim-etal2019,
       author = {{Kim}, Jinho and {Garain}, Sudip K. and {Chakrabarti}, Sandip K. and {Balsara}, Dinshaw S.},
        title = "{General relativistic numerical simulation of sub-Keplerian transonic accretion flows on to rotating black holes: Kerr space-time}",
      journal = {\mnras},
         year = 2019,
        month = jan,
       volume = {482},
       number = {3},
        pages = {3636-3645},
          doi = {10.1093/mnras/sty2953},
archivePrefix = {arXiv},
       eprint = {1810.12469},
 primaryClass = {astro-ph.HE},
       adsurl = {https://ui.adsabs.harvard.edu/abs/2019MNRAS.482.3636K}
}

@ARTICLE{Fishbone-Moncrief1976,
       author = {{Fishbone}, L.~G. and {Moncrief}, V.},
        title = "{Relativistic fluid disks in orbit around Kerr black holes.}",
      journal = {\apj},
         year = 1976,
        month = aug,
       volume = {207},
        pages = {962-976},
          doi = {10.1086/154565},
       adsurl = {https://ui.adsabs.harvard.edu/abs/1976ApJ...207..962F}
}

@ARTICLE{Bondi1952,
       author = {{Bondi}, H.},
        title = "{On spherically symmetrical accretion}",
      journal = {\mnras},
         year = 1952,
        month = jan,
       volume = {112},
        pages = {195},
          doi = {10.1093/mnras/112.2.195},
       adsurl = {https://ui.adsabs.harvard.edu/abs/1952MNRAS.112..195B}
}

@ARTICLE{Ressler-etal2018,
       author = {{Ressler}, S.~M. and {Quataert}, E. and {Stone}, J.~M.},
        title = "{Hydrodynamic simulations of the inner accretion flow of Sagittarius A* fuelled by stellar winds}",
      journal = {\mnras},
         year = 2018,
        month = aug,
       volume = {478},
       number = {3},
        pages = {3544-3563},
          doi = {10.1093/mnras/sty1146},
archivePrefix = {arXiv},
       eprint = {1805.00474},
 primaryClass = {astro-ph.HE},
       adsurl = {https://ui.adsabs.harvard.edu/abs/2018MNRAS.478.3544R}
}

@ARTICLE{Okuda-etal2022,
       author = {{Okuda}, Toru and {Singh}, Chandra B. and {Aktar}, Ramiz},
        title = "{Radiative shock oscillation model for the long-term flares of Sgr A*}",
      journal = {\mnras},
         year = 2022,
        month = aug,
       volume = {514},
       number = {4},
        pages = {5074-5084},
          doi = {10.1093/mnras/stac1630},
archivePrefix = {arXiv},
       eprint = {2206.04919},
 primaryClass = {astro-ph.HE},
       adsurl = {https://ui.adsabs.harvard.edu/abs/2022MNRAS.514.5074O}
}

@ARTICLE{Okuda-etal2019,
       author = {{Okuda}, Toru and {Singh}, Chandra B. and {Das}, Santabrata and {Aktar}, Ramiz and {Nandi}, Anuj and {Dal Pino}, Elisabete M. de Gouveia},
        title = "{A possible model for the long-term flares of Sgr A*}",
      journal = {\pasj},
         year = 2019,
        month = jun,
       volume = {71},
       number = {3},
          eid = {49},
        pages = {49},
          doi = {10.1093/pasj/psz021},
archivePrefix = {arXiv},
       eprint = {1902.02933},
 primaryClass = {astro-ph.HE},
       adsurl = {https://ui.adsabs.harvard.edu/abs/2019PASJ...71...49O}
}

@ARTICLE{Chakrabarti-etal2004,
       author = {{Chakrabarti}, S.~K. and {Acharyya}, K. and {Molteni}, D.},
        title = "{The effect of cooling on time dependent behaviour of accretion flows around black holes}",
      journal = {\aap},
         year = 2004,
        month = jul,
       volume = {421},
        pages = {1-8},
          doi = {10.1051/0004-6361:20034523},
archivePrefix = {arXiv},
       eprint = {astro-ph/0402557},
 primaryClass = {astro-ph},
       adsurl = {https://ui.adsabs.harvard.edu/abs/2004A&A...421....1C}
}

@ARTICLE{Ryu-etal1995,
       author = {{Ryu}, Dongsu and {Brown}, Garry L. and {Ostriker}, Jeremiah P. and {Loeb}, Abraham},
        title = "{Stable and Unstable Accretion Flows with Angular Momentum near a Point Mass}",
      journal = {\apj},
         year = 1995,
        month = oct,
       volume = {452},
        pages = {364},
          doi = {10.1086/176308},
archivePrefix = {arXiv},
       eprint = {astro-ph/9504004},
 primaryClass = {astro-ph},
       adsurl = {https://ui.adsabs.harvard.edu/abs/1995ApJ...452..364R}
}

@ARTICLE{Molteni-etal1994,
       author = {{Molteni}, Diego and {Lanzafame}, Giuseppe and {Chakrabarti}, Sandip K.},
        title = "{Simulation of Thick Accretion Disks with Standing Shocks by Smoothed Particle Hydrodynamics}",
      journal = {\apj},
         year = 1994,
        month = apr,
       volume = {425},
        pages = {161},
          doi = {10.1086/173972},
archivePrefix = {arXiv},
       eprint = {astro-ph/9310047},
 primaryClass = {astro-ph},
       adsurl = {https://ui.adsabs.harvard.edu/abs/1994ApJ...425..161M}
}

@ARTICLE{Molteni-etal1996a,
       author = {{Molteni}, Diego and {Ryu}, Dongsu and {Chakrabarti}, Sandip K.},
        title = "{Numerical Simulations of Standing Shocks in Accretion Flows around Black Holes: A Comparative Study}",
      journal = {\apj},
         year = 1996,
        month = oct,
       volume = {470},
        pages = {460},
          doi = {10.1086/177877},
archivePrefix = {arXiv},
       eprint = {astro-ph/9605116},
 primaryClass = {astro-ph},
       adsurl = {https://ui.adsabs.harvard.edu/abs/1996ApJ...470..460M}
}

@ARTICLE{Molteni-etal1996b,
       author = {{Molteni}, Diego and {Sponholz}, Hanno and {Chakrabarti}, Sandip K.},
        title = "{Resonance Oscillation of Radiative Shock Waves in Accretion Disks around Compact Objects}",
      journal = {\apj},
         year = 1996,
        month = feb,
       volume = {457},
        pages = {805},
          doi = {10.1086/176775},
archivePrefix = {arXiv},
       eprint = {astro-ph/9508022},
 primaryClass = {astro-ph},
       adsurl = {https://ui.adsabs.harvard.edu/abs/1996ApJ...457..805M}
}

@ARTICLE{Lanzafame-etal1998,
       author = {{Lanzafame}, Giuseppe and {Molteni}, D. and {Chakrabarti}, Sandip K.},
        title = "{Smoothed particle hydrodynamic simulations of viscous accretion discs around black holes}",
      journal = {\mnras},
         year = 1998,
        month = sep,
       volume = {299},
       number = {3},
        pages = {799-804},
          doi = {10.1046/j.1365-8711.1998.01816.x},
archivePrefix = {arXiv},
       eprint = {astro-ph/9706248},
 primaryClass = {astro-ph},
       adsurl = {https://ui.adsabs.harvard.edu/abs/1998MNRAS.299..799L}
}

@ARTICLE{Giri-etal2010,
       author = {{Giri}, Kinsuk and {Chakrabarti}, Sandip K. and {Samanta}, Madan M. and {Ryu}, D.},
        title = "{Hydrodynamic simulations of oscillating shock waves in a sub-Keplerian accretion flow around black holes}",
      journal = {\mnras},
         year = 2010,
        month = mar,
       volume = {403},
       number = {1},
        pages = {516-524},
          doi = {10.1111/j.1365-2966.2009.16147.x},
archivePrefix = {arXiv},
       eprint = {0912.1174},
 primaryClass = {astro-ph.HE},
       adsurl = {https://ui.adsabs.harvard.edu/abs/2010MNRAS.403..516G}
}

@ARTICLE{Okuda-Molteni2012,
       author = {{Okuda}, T. and {Molteni}, D.},
        title = "{Low angular momentum flow model for Sgr A*}",
      journal = {\mnras},
         year = 2012,
        month = oct,
       volume = {425},
       number = {4},
        pages = {2413-2421},
          doi = {10.1111/j.1365-2966.2012.21571.x},
archivePrefix = {arXiv},
       eprint = {1206.5338},
 primaryClass = {astro-ph.HE},
       adsurl = {https://ui.adsabs.harvard.edu/abs/2012MNRAS.425.2413O}
}

@ARTICLE{Okuda2014,
       author = {{Okuda}, T.},
        title = "{Low angular momentum flow model II for Sgr A*}",
      journal = {\mnras},
         year = 2014,
        month = jul,
       volume = {441},
       number = {3},
        pages = {2354-2360},
          doi = {10.1093/mnras/stu738},
archivePrefix = {arXiv},
       eprint = {1405.2174},
 primaryClass = {astro-ph.HE},
       adsurl = {https://ui.adsabs.harvard.edu/abs/2014MNRAS.441.2354O}
}

@ARTICLE{Okuda-Das2015,
       author = {{Okuda}, Toru and {Das}, Santabrata},
        title = "{Unstable mass-outflows in geometrically thick accretion flows around black holes}",
      journal = {\mnras},
         year = 2015,
        month = oct,
       volume = {453},
       number = {1},
        pages = {147-156},
          doi = {10.1093/mnras/stv1626},
archivePrefix = {arXiv},
       eprint = {1507.04326},
 primaryClass = {astro-ph.HE},
       adsurl = {https://ui.adsabs.harvard.edu/abs/2015MNRAS.453..147O}
}

@ARTICLE{Palit-etal2019,
       author = {{Palit}, Ishika and {Janiuk}, Agnieszka and {Sukova}, Petra},
        title = "{Effects of adiabatic index on the sonic surface and time variability of low angular momentum accretion flows}",
      journal = {\mnras},
         year = 2019,
        month = jul,
       volume = {487},
       number = {1},
        pages = {755-768},
          doi = {10.1093/mnras/stz1296},
archivePrefix = {arXiv},
       eprint = {1905.02289},
 primaryClass = {astro-ph.HE},
       adsurl = {https://ui.adsabs.harvard.edu/abs/2019MNRAS.487..755P}
}

@ARTICLE{Proga-Begelman2003,
       author = {{Proga}, Daniel and {Begelman}, Mitchell C.},
        title = "{Accretion of Low Angular Momentum Material onto Black Holes: Two-dimensional Hydrodynamical Inviscid Case}",
      journal = {\apj},
         year = 2003,
        month = jan,
       volume = {582},
       number = {1},
        pages = {69-81},
          doi = {10.1086/344537},
archivePrefix = {arXiv},
       eprint = {astro-ph/0208517},
 primaryClass = {astro-ph},
       adsurl = {https://ui.adsabs.harvard.edu/abs/2003ApJ...582...69P}
}

@ARTICLE{Dihingia-etal2024,
author = {{Dihingia}, Indu K. and {Mizuno}, Yosuke},
        title = "{Dynamical Properties of Magnetized Low-angular-momentum Accretion Flows around a Kerr Black Hole}",
      journal = {\apj},
         year = 2024,
        month = may,
       volume = {967},
       number = {1},
          eid = {4},
        pages = {4},
          doi = {10.3847/1538-4357/ad391a},
archivePrefix = {arXiv},
       eprint = {2403.18359},
 primaryClass = {astro-ph.HE},
       adsurl = {https://ui.adsabs.harvard.edu/abs/2024ApJ...967....4D}
}

@ARTICLE{Quataert2004,
       author = {{Quataert}, Eliot},
        title = "{A Dynamical Model for Hot Gas in the Galactic Center}",
      journal = {\apj},
         year = 2004,
        month = sep,
       volume = {613},
       number = {1},
        pages = {322-325},
          doi = {10.1086/422973},
archivePrefix = {arXiv},
       eprint = {astro-ph/0310446},
 primaryClass = {astro-ph},
       adsurl = {https://ui.adsabs.harvard.edu/abs/2004ApJ...613..322Q}
}

@ARTICLE{Cuadra-etal2008,
       author = {{Cuadra}, Jorge and {Nayakshin}, Sergei and {Martins}, Fabrice},
        title = "{Variable accretion and emission from the stellar winds in the Galactic Centre}",
      journal = {\mnras},
         year = 2008,
        month = jan,
       volume = {383},
       number = {2},
        pages = {458-466},
          doi = {10.1111/j.1365-2966.2007.12573.x},
archivePrefix = {arXiv},
       eprint = {0705.0769},
 primaryClass = {astro-ph},
       adsurl = {https://ui.adsabs.harvard.edu/abs/2008MNRAS.383..458C}
}

@article{Chakrabarti:2004uy,
    author = "Chakrabarti, Sandip K. and Das, Santabrata",
    title = "{Properties of accretion shock waves in viscous flows around black holes}",
    eprint = "astro-ph/0402561",
    archivePrefix = "arXiv",
    doi = "10.1111/j.1365-2966.2004.07536.x",
    journal = "Mon. Not. Roy. Astron. Soc.",
    volume = "349",
    pages = "649",
    year = "2004"
}

@ARTICLE{Garain-Kim2023,
       author = {{Garain}, Sudip K. and {Kim}, Jinho},
        title = "{Three-dimensional simulations of advective, sub-Keplerian accretion flow on to non-rotating black holes}",
      journal = {\mnras},
         year = 2023,
        month = mar,
       volume = {519},
       number = {3},
        pages = {4550-4563},
          doi = {10.1093/mnras/stac3736},
archivePrefix = {arXiv},
       eprint = {2212.08310},
 primaryClass = {astro-ph.HE},
       adsurl = {https://ui.adsabs.harvard.edu/abs/2023MNRAS.519.4550G}
}

@ARTICLE{Huang-Singh2025,
       author = {{Huang}, Jun-Xiang and {Singh}, Chandra B.},
        title = "{Relativistic Low Angular Momentum Advective Flows Onto Black Hole and Associated Observational Signatures}",
      journal = {Research in Astronomy and Astrophysics},
         year = 2025,
        month = feb,
       volume = {25},
       number = {2},
          eid = {025013},
        pages = {025013},
          doi = {10.1088/1674-4527/ada2e8},
archivePrefix = {arXiv},
       eprint = {2412.12817},
 primaryClass = {astro-ph.HE},
       adsurl = {https://ui.adsabs.harvard.edu/abs/2025RAA....25b5013H}
}

@article{Dihingia-etal2025,
doi = {10.3847/1538-4357/adf221},
url = {https://dx.doi.org/10.3847/1538-4357/adf221},
year = {2025},
month = {aug},
publisher = {The American Astronomical Society},
volume = {990},
number = {1},
pages = {35},
author = {Dihingia, Indu K. and Uniyal, Akhil and Mizuno, Yosuke},
title = {Imprints of Different Types of Low-angular-momentum Accretion Flow Solutions in General Relativistic Hydrodynamic Simulations},
journal = {The Astrophysical Journal}
}

@ARTICLE{Dihingia-etal2025LA,
       author = {{Dihingia}, Indu K. and {Mizuno}, Yosuke},
        title = "{Emergence of cHz Quasiperiodic Oscillations from a Low-angular-momentum Flow onto a Supermassive Black Hole}",
      journal = {\apjl},
         year = 2025,
        month = mar,
       volume = {982},
       number = {1},
          eid = {L21},
        pages = {L21},
          doi = {10.3847/2041-8213/adbc6d},
archivePrefix = {arXiv},
       eprint = {2503.02337},
 primaryClass = {astro-ph.HE},
       adsurl = {https://ui.adsabs.harvard.edu/abs/2025ApJ...982L..21D}
}

@ARTICLE{Takasao-etal2019,
       author = {{Takasao}, Shinsuke and {Tomida}, Kengo and {Iwasaki}, Kazunari and {Suzuki}, Takeru K.},
        title = "{Giant Protostellar Flares: Accretion-driven Accumulation and Reconnection-driven Ejection of Magnetic Flux in Protostars}",
      journal = {\apjl},
         year = 2019,
        month = jun,
       volume = {878},
       number = {1},
          eid = {L10},
        pages = {L10},
          doi = {10.3847/2041-8213/ab22bb},
archivePrefix = {arXiv},
       eprint = {1902.02007},
 primaryClass = {astro-ph.SR},
       adsurl = {https://ui.adsabs.harvard.edu/abs/2019ApJ...878L..10T}
}

@ARTICLE{Mao-etal2025,
       author = {{Mao}, Jirong and {Dihingia}, Indu K. and {Mizuno}, Yosuke and {Nagataki}, Shigehiro},
        title = "{Low-angular-momentum Black Hole Accretion: First General Relativistic Magnetohydrodynamic Evidence of Standing Shocks}",
      journal = {\apj},
         year = 2025,
        month = sep,
       volume = {990},
       number = {1},
          eid = {12},
        pages = {12},
          doi = {10.3847/1538-4357/adf635},
       adsurl = {https://ui.adsabs.harvard.edu/abs/2025ApJ...990...12M}
}

@article{Belloni:2011vs,
    author = "Belloni, Tomaso M. and Motta, Sara E. and Munoz-Darias, Teodoro",
    title = "{Black hole transients}",
    eprint = "1109.3388",
    archivePrefix = "arXiv",
    primaryClass = "astro-ph.HE",
    journal = "Bull. Astron. Soc. India",
    volume = "39",
    pages = "409",
    year = "2011"
}

@article{Shakura:1972te,
    author = "Shakura, N. I. and Sunyaev, R. A.",
    title = "{Black holes in binary systems. Observational appearance}",
    journal = "Astron. Astrophys.",
    volume = "24",
    pages = "337--355",
    year = "1973"
}

@article{Abramowicz:1988sp,
    author = "Abramowicz, M. A. and Czerny, B. and Lasota, J. P. and Szuszkiewicz, E.",
    title = "{Slim accretion disks}",
    reportNumber = "SISSA-106-87-A",
    doi = "10.1086/166683",
    journal = "Astrophys. J.",
    volume = "332",
    pages = "646",
    year = "1988"
}

@article{Molteni:1996qa,
    author = "Molteni, Diego and Ryu, Dongsu and Chakrabarti, Sandip K.",
    title = "{Numerical simulations of standing shocks in accretion flows around black holes: a comparative study}",
    eprint = "astro-ph/9605116",
    archivePrefix = "arXiv",
    reportNumber = "CNU-AST-96",
    doi = "10.1086/177877",
    journal = "Astrophys. J.",
    volume = "470",
    pages = "460",
    year = "1996"
}

@ARTICLE{Muchotrzebetal1982,
       author = {{Muchotrzeb}, B. and {Paczynski}, B.},
        title = "{Transonic accretion flow in a thin disk around a black hole}",
      journal = {\actaa},
         year = 1982,
        month = jan,
       volume = {32},
       number = {1-2},
        pages = {1-11},
       adsurl = {https://ui.adsabs.harvard.edu/abs/1982AcA....32....1M}
}

@article{Narayan:1994is,
    author = "Narayan, Ramesh and Yi, Insu",
    title = "{Advection dominated accretion: Underfed black holes and neutron stars}",
    eprint = "astro-ph/9411059",
    archivePrefix = "arXiv",
    reportNumber = "CFA-4092",
    doi = "10.1086/176343",
    journal = "Astrophys. J.",
    volume = "452",
    pages = "710",
    year = "1995"
}

@article{Narayan:1994xi,
    author = "Narayan, Ramesh and Yi, In-su",
    title = "{Advection dominated accretion: A Selfsimilar solution}",
    eprint = "astro-ph/9403052",
    archivePrefix = "arXiv",
    reportNumber = "CFA-3809",
    doi = "10.1086/187381",
    journal = "Astrophys. J. Lett.",
    volume = "428",
    pages = "L13",
    year = "1994"
}

@article{Yuan:2014gma,
    author = "Yuan, Feng and Narayan, Ramesh",
    title = "{Hot Accretion Flows Around Black Holes}",
    eprint = "1401.0586",
    archivePrefix = "arXiv",
    primaryClass = "astro-ph.HE",
    doi = "10.1146/annurev-astro-082812-141003",
    journal = "Ann. Rev. Astron. Astrophys.",
    volume = "52",
    pages = "529--588",
    year = "2014"
}

@inproceedings{Novikov:1973kta,
    author = "Novikov, I. D. and Thorne, K. S.",
    title = "{Astrophysics and black holes}",
    booktitle = "{Les Houches Summer School of Theoretical Physics}: {Black Holes}",
    pages = "343--550",
    year = "1973"
}

@ARTICLE{Meliaetal1992,
       author = {{Melia}, Fulvio and {Jokipii}, J.~R. and {Narayanan}, Ajay},
        title = "{A Determination of the Mass of Sagittarius A * from Its Radio Spectral and Source Size Measurements}",
      journal = {\apjl},
         year = 1992,
        month = aug,
       volume = {395},
        pages = {L87},
          doi = {10.1086/186494},
       adsurl = {https://ui.adsabs.harvard.edu/abs/1992ApJ...395L..87M}
}

@article{Narayan:1995ic,
    author = "Narayan, R. and Yi, I. and Mahadevan, R.",
    title = "{Explaining the spectrum of Sagittarius-A* with a model of an accreting black-hole}",
    doi = "10.1038/374623a0",
    journal = "Nature",
    volume = "374",
    pages = "623",
    year = "1995"
}

@article{Ressler:2018yhi,
    author = "Ressler, Sean M. and Quataert, Eliot and Stone, James M.",
    title = "{Hydrodynamic simulations of the inner accretion flow of Sagittarius A* fuelled by stellar winds}",
    eprint = "1805.00474",
    archivePrefix = "arXiv",
    primaryClass = "astro-ph.HE",
    doi = "10.1093/mnras/sty1146",
    journal = "Mon. Not. Roy. Astron. Soc.",
    volume = "478",
    number = "3",
    pages = "3544--3563",
    year = "2018"
}

@article{Fryer:1999mi,
    author = "Fryer, C. L.",
    title = "{Mass limits for black hole formation}",
    eprint = "astro-ph/9902315",
    archivePrefix = "arXiv",
    doi = "10.1086/307647",
    journal = "Astrophys. J.",
    volume = "522",
    pages = "413",
    year = "1999"
}

@ARTICLE{Reesetal1988,
       author = {{Rees}, Martin J.},
        title = "{Tidal disruption of stars by black holes of {}10$^{6}$-{}10$^{8}$ solar masses in nearby galaxies}",
      journal = {\nat},
         year = 1988,
        month = jun,
       volume = {333},
       number = {6173},
        pages = {523-528},
          doi = {10.1038/333523a0},
       adsurl = {https://ui.adsabs.harvard.edu/abs/1988Natur.333..523R}
}

@article{Smith:2001nj,
    author = "Smith, D. M. and Heindl, W. A. and Swank, J. H.",
    title = "{Two different long term behaviors in black hole candidates: evidence for two accretion flows?}",
    eprint = "astro-ph/0103304",
    archivePrefix = "arXiv",
    doi = "10.1086/339167",
    journal = "Astrophys. J.",
    volume = "569",
    pages = "362",
    year = "2002"
}

@article{Tauris:2003pf,
    author = "Tauris, Thomas M. and van den Heuvel, Ed P. J.",
    title = "{Formation and evolution of compact stellar x-ray sources}",
    eprint = "astro-ph/0303456",
    archivePrefix = "arXiv",
    month = "3",
    year = "2003"
}

@BOOK{Chakrabarti1990,
       author = {{Chakrabarti}, Sandip K.},
        title = "{Theory of Transonic Astrophysical Flows}",
         year = 1990,
          doi = {10.1142/1091},
       adsurl = {https://ui.adsabs.harvard.edu/abs/1990ttaf.book.....C}
}

@article{Chakrabarti:1996ns,
    author = "Chakrabarti, Sandip K.",
    title = "{Grand unification of solutions of accretion and winds around black holes and neutron stars}",
    eprint = "astro-ph/9606145",
    archivePrefix = "arXiv",
    doi = "10.1086/177354",
    journal = "Astrophys. J.",
    volume = "464",
    pages = "664",
    year = "1996"
}

@article{Chakrabarti:2019clw,
    author = "Chakrabarti, Sandip K. and Debnath, Dipak and Nagarkoti, Shreeram",
    title = "{Delayed outburst of H 1743{\textendash}322 in 2003 and relation with its other outbursts}",
    eprint = "1902.04833",
    archivePrefix = "arXiv",
    primaryClass = "astro-ph.HE",
    doi = "10.1016/j.asr.2019.02.014",
    journal = "Adv. Space Res.",
    volume = "63",
    pages = "3749--3759",
    year = "2019"
}

@article{Bhowmick:2021lmu,
    author = "Bhowmick, Riya and Debnath, Dipak and Chatterjee, Kaushik and Nagarkoti, Shreeram and Chakrabarti, Sandip Kumar and Sarkar, Ritabrata and Chatterjee, Debjit and Jana, Arghajit",
    title = "{Relation Between Quiescence and Outbursting Properties of GX 339-4}",
    eprint = "2102.02030",
    archivePrefix = "arXiv",
    primaryClass = "astro-ph.HE",
    doi = "10.3847/1538-4357/abe134",
    journal = "Astrophys. J.",
    volume = "910",
    number = "2",
    pages = "138",
    year = "2021"
}

@article{Murchikova:2021rks,
    author = "Murchikova, Lena and Witzel, Gunther",
    title = "{Second-scale Submillimeter Variability of Sagittarius A* during Flaring Activity of 2019: On the Origin of Bright Near-infrared Flares}",
    eprint = "2107.11391",
    archivePrefix = "arXiv",
    primaryClass = "astro-ph.GA",
    doi = "10.3847/2041-8213/ac2308",
    journal = "Astrophys. J. Lett.",
    volume = "920",
    number = "1",
    pages = "L7",
    year = "2021"
}

@article{EventHorizonTelescope:2022urf,
    author = "Akiyama, Kazunori and others",
    collaboration = "Event Horizon Telescope",
    title = "{First Sagittarius A* Event Horizon Telescope Results. V. Testing Astrophysical Models of the Galactic Center Black Hole}",
    eprint = "2311.09478",
    archivePrefix = "arXiv",
    primaryClass = "astro-ph.HE",
    reportNumber = "FERMILAB-PUB-22-419-PPD",
    doi = "10.3847/2041-8213/ac6672",
    journal = "Astrophys. J. Lett.",
    volume = "930",
    number = "2",
    pages = "L16",
    year = "2022"
}

@article{Murchikova:2022aiz,
    author = "Murchikova, Lena and White, Christopher J. and Ressler, Sean M.",
    title = "{Remarkable Correspondence of the Sagittarius A* Submillimeter Variability with a Stellar-wind-fed Accretion Flow Model}",
    eprint = "2204.06170",
    archivePrefix = "arXiv",
    primaryClass = "astro-ph.HE",
    doi = "10.3847/2041-8213/ac75c3",
    journal = "Astrophys. J. Lett.",
    volume = "932",
    number = "2",
    pages = "L21",
    year = "2022"
}

@article{EventHorizonTelescope:2022ago,
    author = "Wielgus, Maciek and others",
    collaboration = "Event Horizon Telescope",
    title = "{Millimeter Light Curves of Sagittarius A* Observed during the 2017 Event Horizon Telescope Campaign}",
    eprint = "2207.06829",
    archivePrefix = "arXiv",
    primaryClass = "astro-ph.HE",
    doi = "10.3847/2041-8213/ac6428",
    journal = "Astrophys. J. Lett.",
    volume = "930",
    number = "2",
    pages = "L19",
    year = "2022"
}

@article{Sponholz:1994ek,
    author = "Sponholz, H. and Molteni, D.",
    title = "{Steady state shocks in accretion disks around a Kerr black hole}",
    eprint = "astro-ph/9407020",
    archivePrefix = "arXiv",
    doi = "10.1093/mnras/271.1.233",
    journal = "Mon. Not. Roy. Astron. Soc.",
    volume = "271",
    pages = "233",
    year = "1994"
}

@article{Chakrabarti:1996ef,
    author = "Chakrabarti, Sandip K.",
    title = "{Global solutions of viscous transonic flows in kerr geometry I: weak viscosity limit}",
    eprint = "astro-ph/9611019",
    archivePrefix = "arXiv",
    doi = "10.1093/mnras/283.1.325",
    journal = "Mon. Not. Roy. Astron. Soc.",
    volume = "283",
    pages = "325",
    year = "1996"
}

@article{Chakrabarti:1996cc,
    author = "Chakrabarti, Sandip K.",
    title = "{Accretion processes on a black hole}",
    eprint = "astro-ph/9605015",
    archivePrefix = "arXiv",
    doi = "10.1016/0370-1573(95)00057-7",
    journal = "Phys. Rept.",
    volume = "266",
    pages = "229--392",
    year = "1996"
}

@article{Komissarov2001,
    author = {Komissarov, S.S.},
    title = {Direct numerical simulations of the Blandford–Znajek effect},
    journal = {Monthly Notices of the Royal Astronomical Society},
    volume = {326},
    number = {3},
    pages = {L41-L44},
    year = {2001},
    month = {09},
    issn = {0035-8711},
    doi = {10.1046/j.1365-8711.2001.04863.x},
    url = {https://doi.org/10.1046/j.1365-8711.2001.04863.x},
    eprint = {https://academic.oup.com/mnras/article-pdf/326/3/L41/2822488/326-3-L41.pdf},
}

@article{Tchekhovskoy:2009ba,
    author = "Tchekhovskoy, Alexander and Narayan, Ramesh and McKinney, Jonathan C.",
    title = "{Black Hole Spin and the Radio Loud/Quiet Dichotomy of Active Galactic Nuclei}",
    eprint = "0911.2228",
    archivePrefix = "arXiv",
    primaryClass = "astro-ph.HE",
    doi = "10.1088/0004-637X/711/1/50",
    journal = "Astrophys. J.",
    volume = "711",
    pages = "50--63",
    year = "2010"
}

@article{Tchekhovskoy:2011zx,
    author = "Tchekhovskoy, Alexander and Narayan, Ramesh and McKinney, Jonathan C.",
    title = "{Efficient Generation of Jets from Magnetically Arrested Accretion on a Rapidly Spinning Black Hole}",
    eprint = "1108.0412",
    archivePrefix = "arXiv",
    primaryClass = "astro-ph.HE",
    doi = "10.1111/j.1745-3933.2011.01147.x",
    journal = "Mon. Not. Roy. Astron. Soc.",
    volume = "418",
    pages = "L79--L83",
    year = "2011"
}

@article{McKinney:2012vh,
    author = "McKinney, Jonathan C. and Tchekhovskoy, Alexander and Blandford, Roger D.",
    title = "{General Relativistic Magnetohydrodynamic Simulations of Magnetically Choked Accretion Flows around Black Holes}",
    eprint = "1201.4163",
    archivePrefix = "arXiv",
    primaryClass = "astro-ph.HE",
    reportNumber = "SLAC-PUB-14950",
    doi = "10.1111/j.1365-2966.2012.21074.x",
    journal = "Mon. Not. Roy. Astron. Soc.",
    volume = "423",
    pages = "3083",
    year = "2012"
}

@article{Tchekhovskoy:2012bg,
    author = "Tchekhovskoy, Alexander and McKinney, Jonathan C.",
    title = "{Prograde and Retrograde Black Holes: Whose Jet is More Powerful?}",
    eprint = "1201.4385",
    archivePrefix = "arXiv",
    primaryClass = "astro-ph.HE",
    doi = "10.1111/j.1745-3933.2012.01256.x",
    journal = "Mon. Not. Roy. Astron. Soc.",
    volume = "423",
    pages = "55",
    year = "2012"
}

@article{Liska:2018ayk,
    author = "Liska, M. and Tchekhovskoy, A. and Ingram, A. and van der Klis, M.",
    title = "{Bardeen{\textendash}Petterson alignment, jets, and magnetic truncation in GRMHD simulations of tilted thin accretion discs}",
    eprint = "1810.00883",
    archivePrefix = "arXiv",
    primaryClass = "astro-ph.HE",
    doi = "10.1093/mnras/stz834",
    journal = "Mon. Not. Roy. Astron. Soc.",
    volume = "487",
    number = "1",
    pages = "550--561",
    year = "2019"
}

@article{Narayan:2003by,
    author = "Narayan, Ramesh and Igumenshchev, Igor V. and Abramowicz, Marek A.",
    title = "{Magnetically arrested disk: an energetically efficient accretion flow}",
    eprint = "astro-ph/0305029",
    archivePrefix = "arXiv",
    doi = "10.1093/pasj/55.6.L69",
    journal = "Publ. Astron. Soc. Jap.",
    volume = "55",
    pages = "L69",
    year = "2003"
}

@article{Balbus:1991ay,
    author = "Balbus, Steven A. and Hawley, John F.",
    title = "{A powerful local shear instability in weakly magnetized disks. 1. Linear analysis. 2. Nonlinear evolution}",
    doi = "10.1086/170270",
    journal = "Astrophys. J.",
    volume = "376",
    pages = "214--233",
    year = "1991"
}

@article{Balbus:1998ja,
    author = "Balbus, Steven A. and Hawley, John F.",
    title = "{Instability, turbulence, and enhanced transport in accretion disks}",
    doi = "10.1103/RevModPhys.70.1",
    journal = "Rev. Mod. Phys.",
    volume = "70",
    pages = "1--53",
    year = "1998"
}

@article{hawley2011assessing,
  title={Assessing quantitative results in accretion simulations: from local to global},
  author={Hawley, John F and Guan, Xiaoyue and Krolik, Julian H},
  journal={The Astrophysical Journal},
  volume={738},
  number={1},
  pages={84},
  year={2011},
  publisher={IOP Publishing}
}

@article{Hawley:2013lga,
    author = "Hawley, John F. and Richers, Sherwood A. and Guan, Xiaoyue and Krolik, Julian H.",
    title = "{Testing Convergence for Global Accretion Disks}",
    eprint = "1306.0243",
    archivePrefix = "arXiv",
    primaryClass = "astro-ph.IM",
    doi = "10.1088/0004-637X/772/2/102",
    journal = "Astrophys. J.",
    volume = "772",
    pages = "102",
    year = "2013"
}

@article{Chakrabarti:1995mx,
    author = "Chakrabarti, Sandip K. and Titarchuk, Lev G.",
    title = "{Spectral properties of accretion disks around galactic and extragalactic black holes}",
    eprint = "astro-ph/9510005",
    archivePrefix = "arXiv",
    reportNumber = "PRINT-95-223 (NASA,GODDARD)",
    doi = "10.1086/176610",
    journal = "Astrophys. J.",
    volume = "455",
    pages = "623",
    year = "1995"
}

@article{Chakrabarti:1993zf,
    author = "Chakrabarti, Sandip K. and Molteni, Diego",
    title = "{Smoothed particle hydrodynamics confronts theory: Formation of standing shocks in accretion disks and winds around black holes}",
    eprint = "astro-ph/9310042",
    archivePrefix = "arXiv",
    doi = "10.1086/173345",
    journal = "Astrophys. J.",
    volume = "417",
    pages = "671",
    year = "1993"
}

@inproceedings{Chakrabarti:1994zq,
    author = "Chakrabarti, Sandip K.",
    title = "{Grand unified model of accretion disks: The SubKeplerian paradigm}",
    booktitle = "{17th Texas Symposium on Relativistic Astrophysics}",
    eprint = "astro-ph/9502040",
    archivePrefix = "arXiv",
    pages = "546--549",
    month = "12",
    year = "1994"
}

@inproceedings{Chakrabarti:1996rc,
    author = "Chakrabarti, Sandip K.",
    title = "{Aspects of accretion processes on a rotating black hole}",
    booktitle = "{18th Conference of the Indian Association for General Relativity and Gravitation}",
    eprint = "astro-ph/9611073",
    archivePrefix = "arXiv",
    reportNumber = "IMSC-117",
    pages = "77--92",
    month = "2",
    year = "1996"
}

@article{Chakrabarti:1997hs,
    author = "Chakrabarti, Sandip K.",
    title = "{Spectral properties of accretion disks around black holes II - sub-keplerian flows with and without shocks}",
    eprint = "astro-ph/9706001",
    archivePrefix = "arXiv",
    doi = "10.1086/304325",
    journal = "Astrophys. J.",
    volume = "484",
    pages = "313",
    year = "1997"
}

@inproceedings{Chakrabarti:1997et,
    author = "Chakrabarti, Sandip K.",
    title = "{Recent progresses of accretion disk models around black holes}",
    booktitle = "{18th Texas Symposium on Relativistic Astrophysics}",
    eprint = "astro-ph/9703051",
    archivePrefix = "arXiv",
    pages = "229--232",
    month = "3",
    year = "1997"
}

@article{Chakrabarti:1998tz,
    author = "Chakrabarti, Sandip K.",
    title = "{Identification of astrophysical black holes}",
    eprint = "astro-ph/9803227",
    archivePrefix = "arXiv",
    journal = "Indian J. Phys. B",
    volume = "72",
    pages = "183",
    year = "1998"
}

@ARTICLE{Fukue:1984,
       author = {{Fukue}, J.},
        title = "{Self-similar transonic flow in the spherically symmetric gravitational field}",
      journal = {\pasj},
         year = 1984,
        month = jan,
       volume = {36},
       number = {1},
        pages = {87-103},
       adsurl = {https://ui.adsabs.harvard.edu/abs/1984PASJ...36...87F}
}

@INPROCEEDINGS{Fukue:1983,
       author = {{Fukue}, J.},
        title = "{Shock Propagations in an Accretion Disk}",
    booktitle = {Theoretical Aspects on Structure, Activity, and Evolution of Galaxies},
         year = 1983,
        month = jan,
        pages = {53},
       adsurl = {https://ui.adsabs.harvard.edu/abs/1983tasa.conf...53F}
}

@ARTICLE{Gaffetetal:1983,
       author = {{Gaffet}, B. and {Fukue}, J.},
        title = "{Strong shock propagation in accretion disks - Self-similar solutions}",
      journal = {\pasj},
         year = 1983,
        month = jan,
       volume = {35},
       number = {3},
        pages = {365-375},
       adsurl = {https://ui.adsabs.harvard.edu/abs/1983PASJ...35..365G}
}

@ARTICLE{Fukue:1983b,
       author = {{Fukue}, J.},
        title = "{Shock propagations in a geometrically thin accretion disk}",
      journal = {\pasj},
         year = 1983,
        month = jan,
       volume = {35},
       number = {3},
        pages = {355-364},
       adsurl = {https://ui.adsabs.harvard.edu/abs/1983PASJ...35..355F}
}

@ARTICLE{Fukue:1983c,
       author = {{Fukue}, J.},
        title = "{Shock propagations in rotating fluid by central explosion Geometrically thick disks}",
      journal = {\pasj},
         year = 1983,
        month = jan,
       volume = {35},
       number = {2},
        pages = {225-233},
       adsurl = {https://ui.adsabs.harvard.edu/abs/1983PASJ...35..225F}
}

@article{Proga:2002ey,
    author = "Proga, Daniel and Begelman, Mitchell C.",
    title = "{Accretion of low angular momentum material onto black holes: 2-D Hydrodynamical inviscid case}",
    eprint = "astro-ph/0208517",
    archivePrefix = "arXiv",
    doi = "10.1086/344537",
    journal = "Astrophys. J.",
    volume = "582",
    pages = "69--81",
    year = "2003"
}

@inproceedings{Czerny:2007pb,
    author = "Czerny, Bozena and Moscibrodzka, Monika and Proga, Daniel and Das, Tapas K. and Siemiginowska, A.",
    title = "{Low angular momentum accretion flow model of Sgr A* activity}",
    booktitle = "{Workshop on Black Holes and Neutron Stars}",
    eprint = "0710.2426",
    archivePrefix = "arXiv",
    primaryClass = "astro-ph",
    pages = "19--30",
    year = "2014"
}

@article{Moscibrodzka:2007fq,
    author = "Moscibrodzka, Monika and Proga, Daniel and Czerny, Bozena and Siemiginowska, Aneta",
    title = "{Accretion of Low Angular Momentum Material onto Black Holes: Radiation Properties of Axisymmetric MHD Flows}",
    eprint = "0707.1403",
    archivePrefix = "arXiv",
    primaryClass = "astro-ph",
    doi = "10.1051/0004-6361:20077703",
    journal = "Astron. Astrophys.",
    volume = "474",
    pages = "1",
    year = "2007"
}

@article{Mach:2018coj,
    author = "Mach, Patryk and Pir{\'o}g, Micha{\l} and Font, Jos{\'e} A.",
    title = "{Relativistic Low Angular Momentum Accretion: Long Time Evolution of Hydrodynamical Inviscid Flows}",
    eprint = "1803.04032",
    archivePrefix = "arXiv",
    primaryClass = "gr-qc",
    doi = "10.1088/1361-6382/aab333",
    journal = "Class. Quant. Grav.",
    volume = "35",
    number = "9",
    pages = "095005",
    year = "2018"
}

@article{Das:2007vf,
    author = "Das, Santabrata and Chakrabarti, Sandip K.",
    editor = "Kleinert, Hagen and Jantzen, Robert T. and Ruffini, Remo",
    title = "{Parameter space study of magnetohydrodynamic flows around magnetized compact objects}",
    eprint = "0706.2896",
    archivePrefix = "arXiv",
    primaryClass = "astro-ph",
    doi = "10.1142/9789812834300_0095",
    journal = "Mon. Not. Roy. Astron. Soc.",
    volume = "374",
    pages = "729",
    year = "2007"
}

@article{Das:2009ie,
    author = "Das, Santabrata and Chakrabarti, Sandip K. and Mondal, Soumen",
    title = "{Studies of dissipative standing shock waves around black holes}",
    eprint = "0909.5513",
    archivePrefix = "arXiv",
    primaryClass = "astro-ph.HE",
    doi = "10.1111/j.1365-2966.2009.15793.x",
    journal = "Mon. Not. Roy. Astron. Soc.",
    volume = "401",
    pages = "2053",
    year = "2010"
}

@article{Uniyal:2023omh,
    author = "Uniyal, Akhil and Chakrabarti, Sayan and Das, Santabrata",
    title = "{Study of relativistic accretion flow in the f(R) theory of gravity}",
    eprint = "2306.14434",
    archivePrefix = "arXiv",
    primaryClass = "astro-ph.HE",
    doi = "10.1016/j.dark.2024.101429",
    journal = "Phys. Dark Univ.",
    volume = "44",
    pages = "101429",
    year = "2024"
}

@article{Uniyal:2024sdv,
    author = "Uniyal, Akhil and Dihingia, Indu K. and Mizuno, Yosuke",
    title = "{A Revisited Equilibrium Solution of the Fishbone and Moncrief Torus for Extended General Relativistic Magnetohydrodynamic Simulations}",
    eprint = "2406.16309",
    archivePrefix = "arXiv",
    primaryClass = "astro-ph.HE",
    doi = "10.3847/1538-4357/ad5b5b",
    journal = "Astrophys. J.",
    volume = "970",
    number = "2",
    pages = "172",
    year = "2024"
}
\bibliographystyle{aasjournalv7}



\end{document}